\documentclass[preprint,12pt]{elsarticle}

\usepackage{graphicx}
\usepackage{multirow}%
\usepackage{amsmath,amssymb,amsfonts}%
\usepackage{mathrsfs}%
\usepackage[title]{appendix}%
\usepackage{xcolor}%
\usepackage{textcomp}%
\usepackage{manyfoot}%
\usepackage{booktabs}%
\usepackage{algorithm}%
\usepackage{algorithmicx}%
\usepackage{algpseudocode}%
\usepackage{listings}%

\usepackage{tcolorbox}
\usepackage{tikz}

\usepackage{amsthm}%

\usepackage[colorlinks=true,citecolor=blue]{hyperref}
\newcommand{\redcite}[1]

\usepackage{etoolbox}
\makeatletter
    \patchcmd{\@author}{\global\let\@fnmark\@empty}{\global\let\@fnmark\@empty\global\let\@corref\@empty}{}{\@latex@error{Failed to patch \string\@author for \string\@corref reset}}
\makeatother

\usetikzlibrary{decorations.pathreplacing,calligraphy}

\newcommand{\srd}{{\sc Signed Roman Domination}}
\newcommand{\sd}{{\sc Signed Domination}}
\newcommand{\rd}{{\sc Roman Domination}}
\newcommand{\ds}{{\sc Dominating Set}}
\newcommand{\rbds}{{\sc Red-Blue Dominating Set}}
\newcommand{\mrss}{{\sc Multidimensional Relaxed Subset Sum}}
\newcommand{\drd}{{\sc Double Roman Domination}}
\newcommand{\erd}{{\sc Edge Roman Domination}}
\newcommand{\wrd}{{\sc Weak Roman Domination}}
\newcommand{\grd}{{\sc Global Roman Domination}}
\newcommand{\trd}{{\sc Total Roman Domination}}

\usepackage{xcolor}
\definecolor{mynicegreen}{RGB}{70,150,70}

\newtheorem{theorem}{Theorem}[section]
\newtheorem{lemma}{Lemma}
{\bfseries}{\itshape}
\newtheorem{claim}{Claim}{\bfseries}{\itshape}{\rmfamily}
\newtheorem{corollary}{Corollary}
\newtheorem{definition}{Definition}{\bfseries}{\itshape}{\rmfamily}

\journal{Theoretical Computer Science}

\begin{document}

\begin{frontmatter}



\title{On the Complexity of Signed Roman Domination}


\author{Sangam Balchandar Reddy}
\ead{21mcpc14@uohyd.ac.in}
\address{School of Computer and Information Sciences, University of Hyderabad, India}

\begin{abstract}
Given a graph $G = (V, E)$, a signed Roman dominating function is a function $f: V \rightarrow \{-1, 1, 2\}$ such that for every vertex $u \in V$: $\sum_{v \in N[u]} f(v) \geq 1$ and for every vertex $u \in V$ with $f(u) = -1$, there exists a vertex $v \in N(u)$ with $f(v) = 2$. The weight of a signed Roman dominating function $f$ is $\sum_{u \in V} f(u)$. The objective of \srd{} (SRD) problem is to compute a signed Roman dominating function with minimum weight. The problem is known to be NP-complete even when restricted to bipartite graphs and planar graphs. In this paper, we advance the complexity study by showing that the problem remains NP-complete on split graphs. In the realm of parameterized complexity, we prove that the problem is W[2]-hard parameterized by weight, even on bipartite graphs. We further show that the problem is W[1]-hard parameterized by feedback vertex set number (and hence also when parameterized by treewidth or clique-width). On the positive side, we present an FPT algorithm parameterized by neighbourhood diversity (and by vertex cover number). Finally, we complement this result by proving that the problem does not admit a polynomial kernel parameterized by vertex cover number unless coNP $\subseteq$ NP/poly.
\end{abstract}




\end{frontmatter}



\section{Introduction}
\ds{} (DS) problem is one of the first few problems that were proved to be NP-complete~\cite{karp} in 1972. Given a graph $G = (V, E)$, a set $S \subseteq V$ is a dominating set if for each vertex $u \in V$, $|N[u] \cap S| \geq 1$. The objective of DS problem is to compute a minimum sized dominating set. Since the inception of DS problem, many variants have been introduced. One such variant is the \rd{} (RD) problem. The concept of RD problem draws inspiration from the strategic positioning of legions to protect the Roman Empire. The RD problem was introduced by Cockayne et al.~\cite{cockayne2004roman} in 2004. Given a graph $G = (V, E)$, a Roman dominating function $f: V \rightarrow{} \{0, 1, 2\}$ is a labeling of vertices such that for each vertex $u \in V$ with $f(u) = 0$, there exists a vertex $v \in N(u)$ with $f(v) = 2$. The goal of RD problem is to obtain a Roman dominating function $f$ with minimum weight. We now define another variant of DS problem, namely, the \sd{} problem (SD). A function $f: V \rightarrow{} \{-1, 1\}$ is a signed dominating function, if $\sum_{v \in N[u]}f(v) \geq 1$, for every $u \in V$. The objective of SD problem is to compute a signed dominating function $f$ with minimum weight.

Various versions of RD problem such as \drd{}~\cite{BEELER201623}, \grd{}~\cite{pushpam2016global}, \erd{}~\cite{pushpam2011weak}, \wrd{}~\cite{cockayne2003secure,henning2003defending} and \trd{}~\cite{liu2013roman} were studied from an algorithmic standpoint. In this work, we consider a relatively new variant, namely the \srd{} problem. The concept of SRD problem was introduced by Ahangar et al.~\cite{ahangar} in 2014. A signed Roman dominating function on a graph $G$ is a function $f : V \rightarrow \{-1,1,2\}$ such that for each $u \in V$: $\sum_{v \in N[u]} f(v) \geq 1$, and every vertex $u$ for which $f(u) = -1$ is adjacent to a vertex $v$ with $f(v) = 2$. The weight of a signed Roman dominating function $f$ is $\sum_{u \in V} f(u)$. The minimum weight over all signed Roman dominating functions for $G$ is denoted by $\gamma_{sR}(G)$. The objective of SRD problem is to compute $\gamma_{sR}(G)$. 
\medskip

\noindent We formally define the problem as follows.
\begin{tcolorbox}
{
\srd{}: \newline
\textit{Input:} An instance $I$ = $(G, k)$, where $G=(V,E)$ is an undirected graph and an integer $k$.\newline
\textit{Output:} Yes, if there exists a function $f: V \rightarrow{} \{-1,1,2\}$ such that \\
(1) $\sum_{u \in V} f(u) \leq k$, \\
(2) $\sum_{v \in N[u]} f(v) \geq 1$ for each vertex $u \in V$ and\\
(3) there exists a vertex $v \in N(u)$ such that $f(v) = 2$ for each vertex $u$ with $f(u) = -1$; No, otherwise.
}
\end{tcolorbox}
\noindent \textbf{Known results.} SRD problem was shown to be NP-complete even when restricted to bipartite graphs and planar graphs~\cite{shao}. There exist methods to obtain $\gamma_{sR}(G)$ in cycles and fan graphs~\cite{behtoei2016signed}. Hong et al.~\cite{hong2020signed} determined $\gamma_{sR}(G)$ for spider graphs and double star graphs.  Yancai et al.~\cite{Yancai} obtained the value of $\gamma_{sR}(G)$ for complete bipartite graphs and wheels. Ahangar et al.~\cite{ahangar} provided several upper and lower bounds on bipartite and general graphs in terms of its order, size and vertex degrees. 
For instance, they have shown that for a graph $G$ with $n$ vertices,
\begin{align*}
    \gamma_{sR}(G) \geq \left(\frac{-2\Delta^2+2\Delta\delta+\Delta+2\delta+3}{(\Delta+1)(2\Delta+\delta+3)}\right) n,
\end{align*}
where $\Delta$ and $\delta$ indicate the maximum degree and minimum degree of the graph $G$, respectively.
There are several combinatorial bounds obtained for the problem on digraphs~\cite{Seyed} and a class of planar graphs called convex polytopes~\cite{Zec2021TheS}.

The DS problem has been extensively studied in the realm of parameterized complexity. Raman et al.~\cite{raman2008short} proved that DS problem is W[2]-complete for split graphs and bipartite graphs. There exists an $\mathcal{O}^*(3^{tw})$-time FPT algorithm for DS problem where $tw$ denotes the treewidth~\cite{van2009dynamic}. 

The parameterized complexity of RD problem has been initiated by Fernau et al.~\cite{henning2}. They showed that RD problem is known to be W[2]-complete for general graphs. They have also presented FPT algorithms for graphs of bounded treewidth and planar graphs. Zheng et al.~\cite{zheng2013kernelization} studied the kernelization aspects of \sd{} on general graphs, planar graphs, $d$-partite graphs, bounded degree graphs, $r$-regular graphs and grid graphs.

Zheng et al.~\cite{zheng2012fpt} initiated the parameterized complexity of SD problem. They showed that the problem is FPT parameterized by treewidth. They also presented kernels for bipartite graphs, bounded degree graphs, regular graphs and planar graphs. Later, Lin et al.~\cite{lin2015algorithms} proved that the problem is W[2]-hard parameterized by weight. They also presented an FPT algorithm for subcubic graphs. 

\medskip \noindent
\textbf{Our results.} 
The primary motivation to study the complexity of SRD problem is due to its ability to combine the properties of both RD problem and SD problem. We attempted to analyze the complexity difference of SRD problem to that of RD problem and SD problem. In this regard, we were able to prove that SRD problem is W[1]-hard parameterized by treewidth whereas both RD problem and SD problem admit FPT algorithms, which is the key contribution of this work.
\medskip

\begin{figure} [t]
\centering
    \begin{tikzpicture} [thick,scale=1.4, every node/.style={scale=0.8}]
        
        \draw[thin, <-] (0, 0.1) -- (0, 0.65); 
        \draw[thin, <-] (0, 1.3) -- (0, 1.95); 
        \draw[thin, <-] (0, 2.35) -- (0, 2.95); 
        \draw[thin, <-] (0, 3.35) -- (0, 4); 

        \draw[thin, <-] (2, 0) -- (2, 0.65); 
        \draw[thin, <-] (2, 1.3) -- (2, 1.9); 
        \draw[thin, <-] (2, 2.55) -- (0.5, 2.95); 
        \draw[thin, <-] (-0.5, 0.1) -- (-2, 0.65); 

        \draw[thin, <-] (-2, 1.4) -- (-2, 1.9); 
        \draw[thin, <-] (-2, 2.35) -- (-0.5, 4); 
        \draw[thin, ->] (1.5, 1.9) -- (0.5, 1.3);

        \draw (0, 4) circle (0cm) node[anchor=south]{Clique-width};
        \draw (0, 3) circle (0cm) node[anchor=south]{Treewidth};
        \draw (0, 2) circle (0cm) node[anchor=south]{Pathwidth};
        \draw (0, 1) circle (0cm) node[anchor=south]{Distance to};
        \draw (0, 0.65) circle (0cm) node[anchor=south]{disjoint paths};
        \draw (0, -0.22) circle (0cm) node[anchor=south]{Vertex cover};
        \draw (0, -0.47) circle (0cm) node[anchor=south]{number};

        \draw (-2, 2) circle (0cm) node[anchor=south]{Modular-width};
        \draw (-2, 1) circle (0cm) node[anchor=south]{Neighbourhood};
        \draw (-2, 0.7) circle (0cm) node[anchor=south]{diversity};
        
        \draw (2, 2.2) circle (0cm) node[anchor=south]{Feedback vertex};
        \draw (2, 1.95) circle (0cm) node[anchor=south]{set number};
        \draw (2, 0.9) circle (0cm) node[anchor=south]{Feedback edge};
        \draw (2, 0.7) circle (0cm) node[anchor=south]{set number};
        \draw (2, -0.35) circle (0cm) node[anchor=south]{Max-leaf number};

        \draw[thin] (-0.8, -0.5) rectangle (0.8, 0.1);
        \draw[thin] (1, -0.4) rectangle (3, 0);
        \draw[thin] (-0.8, 1.3) rectangle (0.8, 0.65);
        \draw[thin] (-0.8, 1.95) rectangle (0.8, 2.35);
        \draw[thin] (-0.8, 2.95) rectangle (0.8, 3.35);
        
        \draw[thin] (-0.8, 4) rectangle (0.8, 4.4);
        \draw[thin] (-2.9, 1.4) rectangle (-1.1, 0.65);
        \draw[thin] (-2.9, 1.9) rectangle (-1.1, 2.35);
        \draw[thin] (1.15, 1.3) rectangle (2.85, 0.65);
        \draw[thin] (1.1, 1.9) rectangle (2.9, 2.55);
        
    \end{tikzpicture}
    \caption{Hasse diagram representing the relation between various structural parameters. A directed edge from a parameter $k_1$ to $k_2$ indicates that $k_1 \leq f(k_2)$ for some computable function $f$.}
    \label{fig: hasse}
\end{figure}
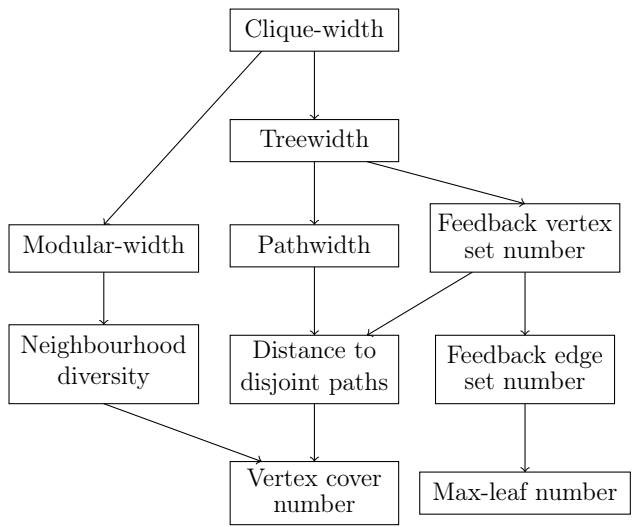

 We obtain the following results for SRD problem in the realm of classical and parameterized complexity. 
 \begin{itemize}
    \item The complexity of SRD problem on split graphs and chordal graphs is open. In this regard, we study the complexity of the problem on split graphs (a subclass of chordal graphs) and prove that the problem is NP-complete.
     \item We initiate the study on parameterized complexity of the problem by proving that the problem is W[2]-hard parameterized by weight even on bipartite graphs.
     \item Both RD problem and SD problem are known to be FPT parameterized by treewidth. In an attempt to solve the complexity of SRD problem parameterized by treewidth, we obtain that the problem is W[1]-hard for the larger parameter feedback vertex set number. This result implies that SRD problem is W[1]-hard parameterized by treewidth and clique-width as well.
     \item As it is known that SRD problem is W[1]-hard parameterized by clique-width, we investigate the parameterized complexity of the problem for a larger parameter neighbourhood diversity and provide an FPT algorithm. 
     \item As the problem is FPT parameterized by vertex cover number, we complement this result by showing that the problem does not admit a polynomial kernel parameterized by vertex cover number unless coNP $\subseteq$ NP/poly.
 \end{itemize}
 The relationship between various graph parameters can be seen in Figure \ref{fig: hasse}.
 
\section{Preliminaries}
\noindent Consider a graph $G = (V, E)$ with $V$ as the vertex set and $E$ as the edge set. For a vertex $u \in V$, the open neighbourhood of $u$ is denoted by $N(u) = \{v: (u,v) \in E\}$ and the closed neighbourhood of $u$ is denoted by $N[u] = N(u) \cup \{u\}$. The open neighbourhood of a set $T \subseteq V$ is denoted by $N(T) = (\bigcup\limits_{u \in T}^{} N(u)) \setminus T$ and the closed neighbourhood by $N[T] = N(T) \cup T$. The degree of a vertex $u$ is represented by $d(u)$ and $d(u) = |N(u)|$. A graph $G$ is $r$-regular if every vertex $u \in V$ has degree $d(u) = r$. A cubic graph is a 3-regular graph. A vertex of degree one is called a pendant vertex. Path length refers to the number of edges in the path. 

Let $f$ be a signed roman dominating function. We use the term \textbf{label} for a vertex $u$ to represent the value that $f$ maps to, i.e., $f(u)$. For a vertex $u$, \textbf{labelSum} is defined as the sum of the labels of all the vertices in its closed neighbourhood, i.e., $\sum_{v \in N[u]} f(v)$. For a set $T$, \textbf{weight} of $T$ is the sum of the labels of all the vertices in the set, i.e., $\sum_{u \in T} f(u)$. We say that a vertex $u \in V$ is \textit{dominated} if it satisfies the signed Roman dominating function constraint. Apart from this, we use the standard notations as defined in~\cite{WEST}.
\medskip

\noindent \textit{Parameterized Complexity.} A problem is said to be fixed-parameter tractable (FPT) with respect to a parameter $k$, if it can be solved by an algorithm with running time $\mathcal{O}(f(k) \cdot n^{\mathcal{O}(1)})$, where $f$ is a computable function and $n$ is the size of the input. We use the notation $\mathcal{O}^*(f(k))$ to suppress polynomial factors in $n$ and represent running times of this form. A problem is W[1]-hard with respect to a parameter if it is believed not to be fixed-parameter tractable. In parameterized complexity, there exists a hierarchy of complexity classes: FPT $\subseteq$ W[1] $\subseteq$ W[2] $\subseteq$ ... $\subseteq$ XP. For more information on \textit{parameterized complexity} and \textit{kernelization}, we refer the reader to~\cite{MC} and~\cite{kernel}, respectively.
\medskip

\noindent We define the graph classes that are used in this paper. 
\begin{definition} [Split graphs]
    A graph $G$ is a split graph if the vertex set $V$ can be partitioned into two sets $V_1$ and $V_2$ such that the subgraph induced on $V_1$ is a clique and the subgraph induced on $V_2$ is an independent set in $G$. 
\end{definition}
 \begin{definition} [Circle graphs]
    A graph is a circle graph if and only if there exists a corresponding circle model in which the chords represent the vertices of the graph and two chords intersect only if the corresponding vertices are adjacent.
\end{definition}
\noindent Now, we define the graph parameters that are used in this paper.
 \begin{definition} [Neighbourhood diversity]
     Let $G = (V, E)$ be a graph. Two vertices $u$ and $v$ are considered to be of the same type if and only if $N(u) \setminus \{v\} = N(v) \setminus \{u\}$. The neighbourhood diversity of $G$ is the smallest value of $t$ for which there exists a partition of $V$ into $t$ sets $V_1, V_2, ..., V_t$ such that all the vertices in each set are of same type.
 \end{definition}
\begin{definition} [Vertex cover number]
    Given a graph $G = (V, E)$, vertex cover number is the cardinality of the smallest set $C \subseteq V$ such that the subgraph induced by $G \setminus C$ is an independent set.
\end{definition}
 \begin{definition} [Feedback vertex set number]
     For a graph $G = (V, E)$, the parameter feedback vertex set number is the cardinality of the smallest vertex subset $F \subseteq V$ such that $G \setminus F$ is a disjoint union of trees.
 \end{definition}
\section{NP-complete on Split Graphs}
In this section, we prove that SRD problem is NP-complete on split graphs, a subclass of chordal graphs. We provide a polynomial-time reduction from DS problem on cubic graphs to prove that SRD problem is NP-complete.

The complexity result related to DS problem on cubic graphs is given as follows.
\begin{theorem}[\cite{Kikuno}]\label{kikunothm}
    DS problem on cubic graphs is NP-complete.
\end{theorem}
Given an instance $I = (G, k)$ of DS problem where $G$ is a cubic graph, we construct an instance $I' = (G', k-3n)$ of SRD problem as follows.
\begin{itemize}
    \item We begin by creating five copies of the vertex set \(V(G)\), denoted by \(A\), $B$, $C$, $D$ and $X$ and include them in \(G'\). Vertex $u_i \in V(G)$ is denoted by $a_i$ in $A$, $b_i$ in $B$, $c_i$ in $C$, $d_i$ in $D$ and $x_i$ in $X$.
    \item We create three sets $E, Y$ and $Z$, each of size $\lceil \frac{2n-k+4}{2} \rceil$. Let $E = \{e_1,e_2,...,e_{\lceil \frac{2n-k+4}{2} \rceil}\}$, $Y = \{y_1,y_2,...,y_{\lceil \frac{2n-k+4}{2} \rceil}\}$ and $Z = \{z_1,z_2,...,z_{\lceil \frac{2n-k+4}{2} \rceil}\}$.
    \item For each pair of vertices \( a_i \in A \) and \( x_j \in X \), we add an edge between \( a_i \) and \( x_j \) if and only if \( u_i \in N_G(u_j) \).
    \item For each $i \in [n]$, we make $x_i \in X$ adjacent to $a_i, b_i$, $c_i$ and $d_i$.
    \item For each $i \in [\lceil \frac{2n-k+4}{2} \rceil]$, we make $e_i \in E$ adjacent to both $y_i$ and $z_i$.
    \item We form a clique from the vertices of the set $A \cup B \cup C \cup D \cup E$.
\end{itemize}
\begin{figure}
    \centering
    \begin{tikzpicture} [scale = 0.8]
        \filldraw[black] (0, 0) circle (3pt);
        \filldraw[black] (2, 1) circle (3pt);
        \filldraw[black] (2, -1) circle (3pt);
        \filldraw[black](4, 1) circle (3pt);
        \filldraw[black] (4, -1) circle (3pt);
        \filldraw[black] (6, 0) circle (3pt);
        \draw[thin] (0.1, 0.1) -- (1.9, 0.9);
        \draw[thin] (0.1, -0.1) -- (1.9, -0.9);
        \draw[thin] (0.15, 0) -- (5.85, 0);
        \draw[thin] (2, 0.85) -- (2, -0.85);
        \draw[thin] (4, 0.85) -- (4, -0.85);
        \draw[thin] (2.15, 1) -- (3.85, 1);
        \draw[thin] (2.15, -1) -- (3.85, -1);
        \draw[thin] (4.1, 0.9) -- (5.9, 0.1);
        \draw[thin] (4.1, -0.9) -- (5.9, -0.1);
        \filldraw (0, 0.1) circle (0cm) node[anchor=south]{$v_1$};
        \filldraw (2, 1.1) circle (0cm) node[anchor=south]{$v_2$};
        \filldraw (2, -1.8) circle (0cm) node[anchor=south]{$v_3$};
        \filldraw (4, 1.1) circle (0cm) node[anchor=south]{$v_4$};
        \filldraw (4, -1.8) circle (0cm) node[anchor=south]{$v_5$};
        \filldraw (6, 0.1) circle (0cm) node[anchor=south]{$v_6$};
        \filldraw (6, -5) circle (0cm) node[anchor=south]{};
        \filldraw (8, 0) circle (0cm) node[anchor=south]{$G$};
    \end{tikzpicture}
     \begin{tikzpicture} [thick,scale=1.5, every node/.style={scale=0.8}]
        \draw (0, -1.5) ellipse (1.1 and 4);
        \draw (4, -1.8) ellipse (0.8 and 2.8);
        \draw[thin, gray] (-0.93, 0.6) -- (0.93, 0.6);
        \draw[thin, gray] (-1.07, -0.7) -- (1.07, -0.7);
        \draw[thin, gray] (-1.08, -2) -- (1.08, -2);
        \draw[thin, gray] (-1, -3.25) -- (1, -3.25);
        \draw[thin, gray] (3.25, -3.05) -- (4.75, -3.05);
        \draw[thin, gray] (3.2, -1.6) -- (4.8, -1.6);

        \draw[blue] (4, -0.1) -- (0, 0.4);
        \draw[thin, gray] (4, -0.3) -- (0, 0.225);
        \draw[thin, gray] (4, -0.7) -- (0, -0.125);
        \draw[thin, gray] (4, -0.9) -- (0, -0.3);
        \draw[thin, gray] (4, -0.5) -- (0, 0.05);
        \draw[thin, gray] (4, -1.1) -- (0, -0.475);

        \draw[blue] (4, -0.1) -- (0, -0.875);
        \draw[thin, gray] (4, -0.3) -- (0, -1.05);
        \draw[thin, gray] (4, -0.5) -- (0, -1.225);
        \draw[thin, gray] (4, -0.7) -- (0, -1.4);
        \draw[thin, gray] (4, -0.9) -- (0, -1.575);
        \draw[thin, gray] (4, -1.1) -- (0, -1.75);

        \draw[blue] (4, -0.1) -- (0, -2.175);
        \draw[thin, gray] (4, -0.3) -- (0, -2.35);
        \draw[thin, gray] (4, -0.5) -- (0, -2.525);
        \draw[thin, gray] (4, -0.7) -- (0, -2.7);
        \draw[thin, gray] (4, -0.9) -- (0, -2.875);
        \draw[thin, gray] (4, -1.1) -- (0, -3.05);

        \draw[blue] (4, -0.1) -- (0, 2.3);
        \draw[thin, gray] (4, -0.3) -- (0, 2.3);
        \draw[thin, gray] (4, -0.5) -- (0, 2.3);
        \draw[thin, gray] (4, -1.1) -- (0, 2.3);
        \draw[thin, gray] (4, -0.3) -- (0, 2);
        \draw[blue] (4, -0.1) -- (0, 2);
        \draw[thin, gray] (4, -0.5) -- (0, 2);
        \draw[thin, gray] (4, -0.7) -- (0, 2);
        \draw[thin, gray] (4, -0.5) -- (0, 1.7);
        \draw[blue] (4, -0.1) -- (0, 1.7);
        \draw[thin, gray] (4, -0.3) -- (0, 1.7);
        \draw[thin, gray] (4, -0.7) -- (0, 1.7);
        \draw[thin, gray] (4, -0.7) -- (0, 1.4);
        \draw[thin, gray] (4, -0.3) -- (0, 1.4);
        \draw[thin, gray] (4, -1.1) -- (0, 1.4);
        \draw[thin, gray] (4, -1.1) -- (0, 1.4);
        \draw[thin, gray] (4, -0.9) -- (0, 1.1);
        \draw[thin, gray] (4, -0.5) -- (0, 1.1);
        \draw[thin, gray] (4, -0.7) -- (0, 1.1);
        \draw[thin, gray] (4, -1.1) -- (0, 1.1);
        \draw[thin, gray] (4, -1.1) -- (0, 0.8);
        \draw[blue] (4, -0.1) -- (0, 0.8);
        \draw[thin, gray] (4, -1.1) -- (0, 0.8); 
        \draw[thin, gray] (4, -0.9) -- (0, 0.8);    

        \draw[thin, gray] (0, -3.5) -- (4, -1.85);
        \draw[thin, gray] (0, -3.5) -- (4, -3.35);        
        \draw[thin, gray] (0, -3.8) -- (4, -2);
        \draw[thin, gray] (0, -3.8) -- (4, -3.5);        
        \draw[thin, gray] (0, -4.1) -- (4, -2.15);
        \draw[thin, gray] (0, -4.1) -- (4, -3.65);        
        \draw[thin, gray] (0, -4.4) -- (4, -2.3);
        \draw[thin, gray] (0, -4.4) -- (4, -3.8);      
        \draw[thin, gray] (0, -4.7) -- (4, -2.45);
        \draw[thin, gray] (0, -4.7) -- (4, -3.95);        
        \draw[thin, gray] (0, -5) -- (4, -2.6);
        \draw[thin, gray] (0, -5) -- (4, -4.1);        
        \draw[thin, gray] (0, -5.3) -- (4, -2.75);
        \draw[thin, gray] (0, -5.3) -- (4, -4.25);

        \filldraw (0, 1.7) circle (1pt) node[anchor=south]{};
        \filldraw (0, 1.1) circle (1pt) node[anchor=south]{};
        \filldraw (0, 1.4) circle (1pt) node[anchor=south]{};
        \filldraw (0, 0.8) circle (1pt) node[anchor=south]{};
        \filldraw (0, 2) circle (1pt) node[anchor=south]{};
        \filldraw (0, 2.3) circle (1pt) node[anchor=south]{};

        \filldraw (-0.2, 1.55) circle (0cm) node[anchor=south]{$a_3$};
        \filldraw (-0.2, 1.25) circle (0cm) node[anchor=south]{$a_4$};
        \filldraw (-0.2, 0.95) circle (0cm) node[anchor=south]{$a_5$};
        \filldraw (-0.2, 0.65) circle (0cm) node[anchor=south]{$a_6$}; 
        \filldraw (-0.2, 1.85) circle (0cm) node[anchor=south]{$a_2$};
        \filldraw (-0.2, 2.15) circle (0cm) node[anchor=south]{$a_1$};  

        \filldraw (4.25, -0.25) circle (0cm) node[anchor=south]{$x_1$};
        \filldraw (4.25, -0.45) circle (0cm) node[anchor=south]{$x_2$};
        \filldraw (4.25, -0.65) circle (0cm) node[anchor=south]{$x_3$};
        \filldraw (4.25, -0.85) circle (0cm) node[anchor=south]{$x_4$};
        \filldraw (4.25, -1.05) circle (0cm) node[anchor=south]{$x_5$};
        \filldraw (4.25, -1.25) circle (0cm) node[anchor=south]{$x_6$};

        \filldraw (-0.25, 0.25) circle (0cm) node[anchor=south]{$b_1$};
        \filldraw (-0.25, 0.05) circle (0cm) node[anchor=south]{$b_2$};
        \filldraw (-0.25, -0.125) circle (0cm) node[anchor=south]{$b_3$};
        \filldraw (-0.25, -0.3) circle (0cm) node[anchor=south]{$b_4$};
        \filldraw (-0.25, -0.5) circle (0cm) node[anchor=south]{$b_5$};
        \filldraw (-0.25, -0.7) circle (0cm) node[anchor=south]{$b_6$};     

        \filldraw (-0.25, -1) circle (0cm) node[anchor=south]{$c_1$};
        \filldraw (-0.25, -1.2) circle (0cm) node[anchor=south]{$c_2$};
        \filldraw (-0.25, -1.35) circle (0cm) node[anchor=south]{$c_3$};
        \filldraw (-0.25, -1.55) circle (0cm) node[anchor=south]{$c_4$};
        \filldraw (-0.25, -1.7) circle (0cm) node[anchor=south]{$c_5$};
        \filldraw (-0.25, -1.925) circle (0cm) node[anchor=south]{$c_6$}; 

        \filldraw (-0.25, -2.3) circle (0cm) node[anchor=south]{$d_1$};
        \filldraw (-0.25, -2.5) circle (0cm) node[anchor=south]{$d_2$};
        \filldraw (-0.25, -2.675) circle (0cm) node[anchor=south]{$d_3$};
        \filldraw (-0.25, -2.85) circle (0cm) node[anchor=south]{$d_4$};
        \filldraw (-0.25, -3.05) circle (0cm) node[anchor=south]{$d_5$};
        \filldraw (-0.25, -3.25) circle (0cm) node[anchor=south]{$d_6$}; 

        \filldraw (-0.2, -3.625) circle (0cm) node[anchor=south]{$e_1$};
        \filldraw (-0.2, -3.9) circle (0cm) node[anchor=south]{$e_2$};
        \filldraw (-0.2, -4.2) circle (0cm) node[anchor=south]{$e_3$};
        \filldraw (-0.2, -4.5) circle (0cm) node[anchor=south]{$e_4$};
        \filldraw (-0.2, -4.8) circle (0cm) node[anchor=south]{$e_5$};
        \filldraw (-0.2, -5.1) circle (0cm) node[anchor=south]{$e_6$};
        \filldraw (-0.2, -5.4) circle (0cm) node[anchor=south]{$e_7$};
        
        \filldraw (4.25, -1.9) circle (0cm) node[anchor=south]{$y_1$};
        \filldraw (4.25, -2.08) circle (0cm) node[anchor=south]{$y_2$};
        \filldraw (4.25, -2.26) circle (0cm) node[anchor=south]{$y_3$};
        \filldraw (4.25, -2.44) circle (0cm) node[anchor=south]{$y_4$};
        \filldraw (4.25, -2.62) circle (0cm) node[anchor=south]{$y_5$};
        \filldraw (4.25, -2.8) circle (0cm) node[anchor=south]{$y_6$};
        \filldraw (4.25, -2.98) circle (0cm) node[anchor=south]{$y_7$};
        
        \filldraw (4.2, -3.4) circle (0cm) node[anchor=south]{$z_1$};
        \filldraw (4.2, -3.58) circle (0cm) node[anchor=south]{$z_2$};
        \filldraw (4.2, -3.76) circle (0cm) node[anchor=south]{$z_3$};
        \filldraw (4.2, -3.94) circle (0cm) node[anchor=south]{$z_4$};
        \filldraw (4.2, -4.12) circle (0cm) node[anchor=south]{$z_5$};
        \filldraw (4.2, -4.3) circle (0cm) node[anchor=south]{$z_6$};
        \filldraw (4.2, -4.48) circle (0cm) node[anchor=south]{$z_7$};
        
        \filldraw (0, 0.4) circle (1pt) node[anchor=south]{};
        \filldraw (0, 0.225) circle (1pt) node[anchor=south]{};
        \filldraw (0, 0.05) circle (1pt) node[anchor=south]{};
        \filldraw (0, -0.125) circle (1pt) node[anchor=south]{};
        \filldraw (0, -0.3) circle (1pt) node[anchor=south]{};
        \filldraw (0, -0.475) circle (1pt) node[anchor=south]{};

        \filldraw (0, -0.875) circle (1pt) node[anchor=south]{};
        \filldraw (0, -1.05) circle (1pt) node[anchor=south]{};
        \filldraw (0, -1.225) circle (1pt) node[anchor=south]{};
        \filldraw (0, -1.4) circle (1pt) node[anchor=south]{};
        \filldraw (0, -1.575) circle (1pt) node[anchor=south]{};
        \filldraw (0, -1.75) circle (1pt) node[anchor=south]{};

        \filldraw (0, -2.175) circle (1pt) node[anchor=south]{};
        \filldraw (0, -2.35) circle (1pt) node[anchor=south]{};
        \filldraw (0, -2.525) circle (1pt) node[anchor=south]{};
        \filldraw (0, -2.7) circle (1pt) node[anchor=south]{};
        \filldraw (0, -2.875) circle (1pt) node[anchor=south]{};
        \filldraw (0, -3.05) circle (1pt) node[anchor=south]{};

        \filldraw (0, -3.5) circle (1pt) node[anchor=south]{};
        \filldraw (0, -3.8) circle (1pt) node[anchor=south]{};
        \filldraw (0, -4.1) circle (1pt) node[anchor=south]{};
        \filldraw (0, -4.4) circle (1pt) node[anchor=south]{};
        \filldraw (0, -4.7) circle (1pt) node[anchor=south]{};
        \filldraw (0, -5) circle (1pt) node[anchor=south]{};
        \filldraw (0, -5.3) circle (1pt) node[anchor=south]{};

        \filldraw (4, -0.1) circle (1pt) node[anchor=south]{};
        \filldraw (4, -0.3) circle (1pt) node[anchor=south]{};
        \filldraw (4, -0.5) circle (1pt) node[anchor=south]{};
        \filldraw (4, -0.7) circle (1pt) node[anchor=south]{};
        \filldraw (4, -0.9) circle (1pt) node[anchor=south]{};
        \filldraw (4, -1.1) circle (1pt) node[anchor=south]{};
        
        \filldraw (4, -1.85) circle (1pt) node[anchor=south]{};
        \filldraw (4, -2) circle (1pt) node[anchor=south]{};
        \filldraw (4, -2.15) circle (1pt) node[anchor=south]{};
        \filldraw (4, -2.3) circle (1pt) node[anchor=south]{};
        \filldraw (4, -2.45) circle (1pt) node[anchor=south]{};
        \filldraw (4, -2.6) circle (1pt) node[anchor=south]{};
        \filldraw (4, -2.75) circle (1pt) node[anchor=south]{};

        \filldraw (4, -3.35) circle (1pt) node[anchor=south]{};
        \filldraw (4, -3.5) circle (1pt) node[anchor=south]{};
        \filldraw (4, -3.65) circle (1pt) node[anchor=south]{};
        \filldraw (4, -3.8) circle (1pt) node[anchor=south]{};
        \filldraw (4, -3.95) circle (1pt) node[anchor=south]{};
        \filldraw (4, -4.1) circle (1pt) node[anchor=south]{};
        \filldraw (4, -4.25) circle (1pt) node[anchor=south]{};
        
        \filldraw (-1.1, 1.2) circle (0cm) node[anchor=south]{$A$};
        \filldraw (-1.3, -0.2) circle (0cm) node[anchor=south]{$B$};
        \filldraw (-1.4, -1.6) circle (0cm) node[anchor=south]{$C$};
        \filldraw (-1.35, -2.8) circle (0cm) node[anchor=south]{$D$};
        \filldraw (-1.2, -4.2) circle (0cm) node[anchor=south]{$E$};
        \filldraw (0.1, -6) circle (0cm) node[anchor=south]{clique};
        \filldraw (4.2, -5.1) circle (0cm) node[anchor=south]{independent set};
        \filldraw (5.05, -2.6) circle (0cm) node[anchor=south]{$Y$};
        \filldraw (5, -0.5) circle (0cm) node[anchor=south]{$X$};
        \filldraw (4.8, -4) circle (0cm) node[anchor=south]{$Z$};
        \filldraw (6.5, -2) circle (0cm) node[anchor=south]{$G'$};
     \end{tikzpicture}
    \caption{Reduced instance $G'$ from an instance of the dominating set $G$ and $k = 2$. Edges between the vertices of the clique $A \cup B \cup C \cup D \cup E$ are not shown in the figure. The edges incident on the vertex $x_1$ are colored in blue.}
    \label{fig:fig2}
\end{figure}
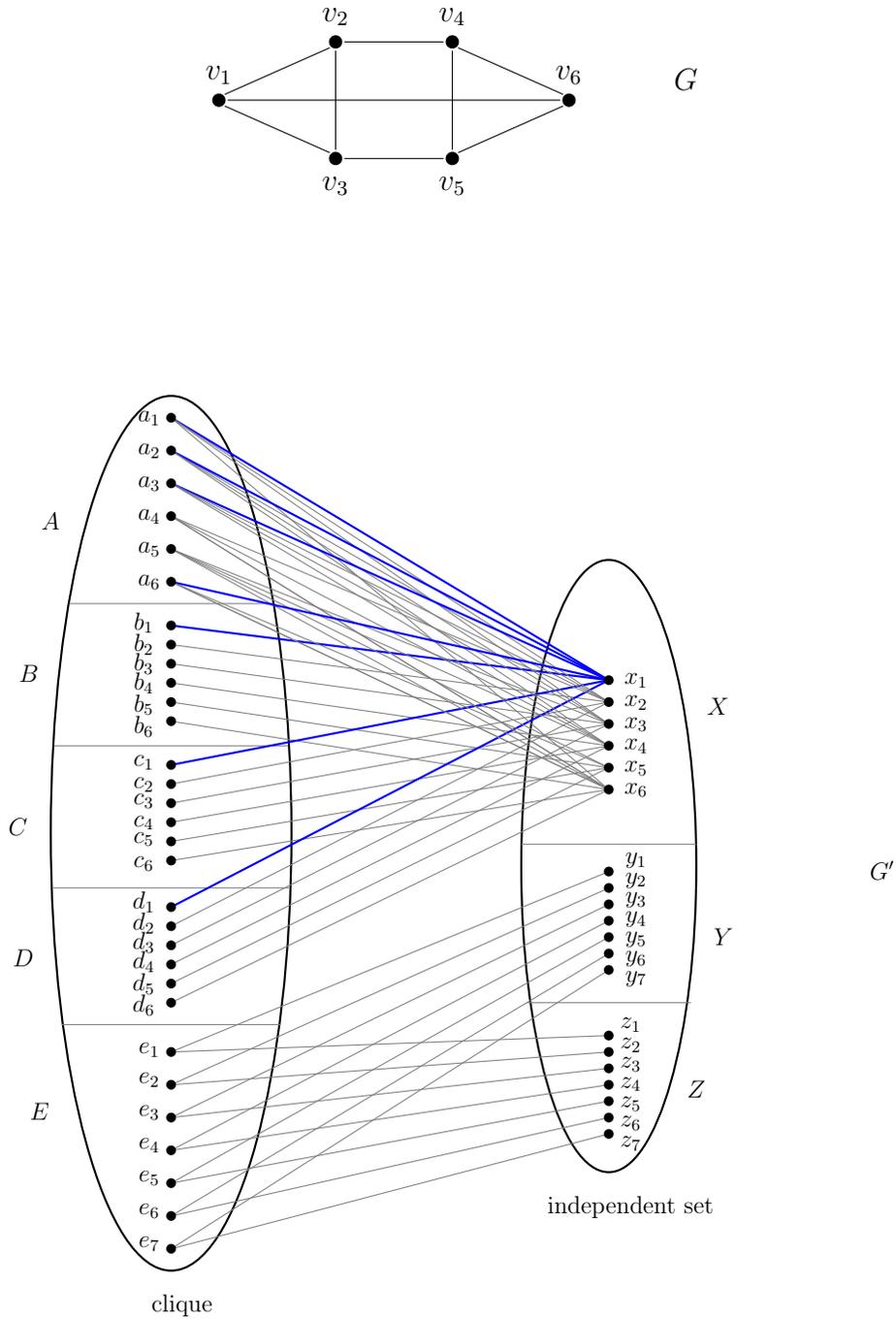
This concludes the construction of $G'$. See Figure \ref{fig:fig2} for an illustration.

Observe that $G'$ is a split graph, where the subgraph induced by $A \cup B \cup C \cup D \cup E$ forms a clique and the subgraph induced by $X \cup Y \cup Z$ forms an independent set in $G'$.
\begin{lemma}~\label{lemma1}
    $I$ has a dominating set of size at most $k$ if and only if $I'$ has a signed Roman dominating function of weight at most $k-3n$.
\end{lemma}
\begin{proof}
$(\Rightarrow)$ Let $S$ be a dominating set of size at most $k$. We obtain a signed Roman dominating function $f$ as follows.
\[
f(w) = 
\begin{cases}
-1, & \text{if } w \in \left(B \cup C \cup D \cup X \cup Y \cup Z \right), \\
1, & \text{if } w \in \left(\bigcup\limits_{u_i \notin S} \{a_i\} \right), \\
2, & \text{if } w\in \left( E \cup \bigcup\limits_{u_i \in S} \{a_i\} \right)
\end{cases}
\]
\begin{itemize}
    \item Vertex $u \in X$ is adjacent to four vertices in $A$, at least one with the label 2 while the others are assigned the label 1. Vertex $u$ is also adjacent to three vertices with the label $-1$, one from each of the sets $B$, $C$ and $D$. Although $u$ has the label $-1$, the labelSum of $u$ is at least one. Hence, we have that, each vertex $u \in X$ has a labelSum of at least one and also has a neighbour in $A$ with the label 2.
    \item Vertex $u \in Y \cup Z$ has the label $-1$ and the only neighbour of $u$ in $E$ has the label 2. Therefore, we have that each vertex $u \in Y \cup Z$ has a labelSum of one and also has a neighbour in $E$ with the label 2.
    \item The weight of the clique $A \cup B \cup C \cup D \cup E$ is at least four.
    \item Vertex $u \in A$ is adjacent to all the vertices in the clique  $A \cup B \cup C \cup D \cup E$ and also adjacent to three vertices in $X$ with the label $-1$. We have that each vertex $u \in A$ has a labelSum of at least one and also has a neighbour in $A$ with the label 2.
    \item Vertex $u \in E$ is adjacent to all the vertices in the clique  $A \cup B \cup C \cup D \cup E$ and also adjacent to two vertices in $Y \cup Z$ with the label $-1$. We have that each vertex $u \in E$ has a labelSum of at least one.
    \item Vertex $u \in B \cup C \cup D$ is adjacent to all the vertices in the clique  $A \cup B \cup C \cup D \cup E$ and also adjacent to one vertex in $X$ with the label $-1$. We have that each vertex $u \in B \cup C \cup D$ has a labelSum of at least three and also has a neighbour in $E$ with the label 2.
\end{itemize}
The weight of $A$ is $n+k$, $B$ is $-n$, $C$ is $-n$, $D$ is $-n$, $X$ is $-n$ and $E \cup Y \cup Z$ is zero. As each vertex in $G'$ is dominated, we conclude that $f$ is a signed Roman dominating function of weight $k-3n$. 
\medskip

\noindent $(\Leftarrow)$ Let $f$ be a signed Roman dominating function of weight at most $k-3n$. We obtain the dominating set $S$ as follows. 
\begin{itemize}
    \item Each vertex in $E$ is adjacent to two pendant vertices in $Y \cup Z$. We label each vertex in $E$ with 2 and each vertex in $Y \cup Z$ with $-1$, leading to a weight of zero for the set $E \cup Y \cup Z$. Any other labeling for the vertices in $E \cup Y \cup Z$ would lead to a positive weight.
    \item At this point, we have $5n$ more vertices to label with a weight of $k-3n$. In order to have a positive labelSum for the vertices in $A$, the weight of $A \cup B \cup C \cup D$ should be at least $k-2n-1$. Hence, we must label all the vertices in $X$ with $-1$.
    \item Each vertex in $X$ is adjacent to three vertices in $A$ and exactly one vertex in each of the three sets $B$, $C$ and $D$. Therefore, the vertices in $B \cup C \cup D$ are assigned the label $-1$ in $f$.
    \item We have a weight of $n+k$ for the vertices in $A$. At most $k$ vertices in $A$ can be labeled 2 while the others are labeled 1. For each vertex in $X$ to have a positive labelSum, at least one of its neighbours in $A$ must be labeled 2.
    \item $k$ vertices from $A$ that are assigned the label 2 will correspond to the $k$-vertex dominating set $S$ for $G$.
\end{itemize}   
\end{proof}
Hence, from Theorem \ref{kikunothm} and Lemma \ref{lemma1}, we arrive at the following theorem.
\begin{theorem}
    SRD problem on split graphs is NP-complete.
\end{theorem}
As split graphs is a subclass of chordal graphs, we have the following result.
\begin{corollary}
    SRD problem on chordal graphs is NP-complete.
\end{corollary}

\section{W[2]-hard Parameterized by Weight}
In this section, we investigate the parameterized complexity of SRD problem parameterized by weight. We prove that the problem is W[2]-hard even on bipartite graphs. To prove this, we provide a parameterized reduction from DS problem. Given a graph $G = (V, E)$, a set $S \subseteq V$ is a dominating set if every vertex $u \in V\setminus S$ has a neighbour in $S$. The decision version of DS problem asks whether there exists a dominating set of size at most $k$.

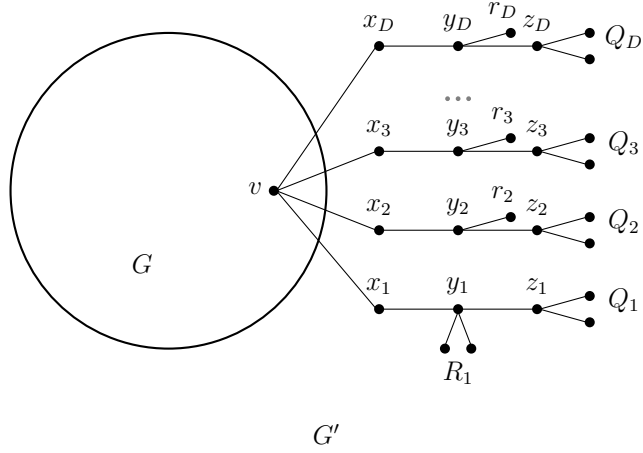
\begin{figure} [t]
\centering
    \begin{tikzpicture} [thick,scale=0.7, every node/.style={scale=0.85}]
        \draw (5, 5.25) circle (3);
        
        \filldraw (7, 5.25) circle (0.075cm);
        
        \filldraw (9, 8) circle (0.075cm);
        \filldraw (10.5, 8) circle (0.075cm);
        \filldraw (12, 8) circle (0.075cm);

        \filldraw (9, 4.5) circle (0.075cm);
        \filldraw (10.5, 4.5) circle (0.075cm);
        \filldraw (12, 4.5) circle (0.075cm);
                
        \filldraw (9, 6) circle (0.075cm);
        \filldraw (10.5, 6) circle (0.075cm);
        \filldraw (12, 6) circle (0.075cm);

        \filldraw (9, 3) circle (0.075cm);
        \filldraw (10.5, 3) circle (0.075cm);
        \filldraw (12, 3) circle (0.075cm);

        \filldraw (10.25, 2.25) circle (0.075cm);
        \filldraw (10.75, 2.25) circle (0.075cm);

        \filldraw (13, 3.25) circle (0.075cm);
        \filldraw (13, 2.75) circle (0.075cm);

        \filldraw (13, 7.75) circle (0.075cm);
        \filldraw (13, 8.25) circle (0.075cm);
               
        \filldraw (13, 4.25) circle (0.075cm);
        \filldraw (13, 4.75) circle (0.075cm);

        \filldraw (13, 5.75) circle (0.075cm);
        \filldraw (13, 6.25) circle (0.075cm);

        \draw[thin] (7.05, 5.25) -- (8.95, 8);
        \draw[thin] (7.05, 5.25) -- (8.95, 4.5);
        \draw[thin] (7.05, 5.25) -- (8.95, 3);
        \draw[thin] (7.05, 5.25) -- (8.95, 6);

        \draw[thin] (10.45, 8) -- (9.05, 8);
        \draw[thin] (10.45, 4.5) -- (9.05, 4.5);
        \draw[thin] (10.45, 6) -- (9.05, 6);
        \draw[thin] (10.45, 3) -- (9.05, 3);

        \draw[thin] (10.55, 8) -- (11.95, 8);
        \draw[thin] (10.55, 4.5) -- (11.95, 4.5);
        \draw[thin] (10.55, 6) -- (11.95, 6);
        \draw[thin] (10.55, 3) -- (11.95, 3);

        \draw[thin] (10.5, 2.95) -- (10.25, 2.3);
        \draw[thin] (10.5, 2.95) -- (10.75, 2.3);

        \draw[thin] (12.05, 3) -- (12.95, 2.75);
        \draw[thin] (12.05, 3) -- (12.95, 3.25);

        \draw[thin] (12.05, 8) -- (12.95, 7.75);
        \draw[thin] (12.05, 8) -- (12.95, 8.25);

        \draw[thin] (12.05, 4.5) -- (12.95, 4.25);
        \draw[thin] (12.05, 4.5) -- (12.95, 4.75);

        \draw[thin] (12.05, 6) -- (12.95, 5.75);
        \draw[thin] (12.05, 6) -- (12.95, 6.25);

        \draw[thin] (10.55, 4.5) -- (11.45, 4.75);
        \draw[thin] (10.55, 6) -- (11.45, 6.25);
        \draw[thin] (10.55, 8) -- (11.45, 8.25);

        \filldraw (11.5, 4.75) circle (0.075cm);
        \filldraw (11.5, 6.25) circle (0.075cm);
        \filldraw (11.5, 8.25) circle (0.075cm);

        \draw (10.3, 7)[gray] circle (0.025cm);
        \draw (10.5, 7)[gray] circle (0.025cm);
        \draw (10.7, 7)[gray] circle (0.025cm);

        \draw (6.65, 5) circle (0cm) node[anchor=south]{$v$};

        \draw (9, 3.1) circle (0cm) node[anchor=south]{$x_1$};
        \draw (10.5, 3.1) circle (0cm) node[anchor=south]{$y_1$};
        \draw (10.5, 1.4) circle (0cm) node[anchor=south]{$R_1$};
        \draw (12, 3.1) circle (0cm) node[anchor=south]{$z_1$};

        \draw (9, 4.6) circle (0cm) node[anchor=south]{$x_2$};
        \draw (10.5, 4.6) circle (0cm) node[anchor=south]{$y_2$};
        \draw (12, 4.6) circle (0cm) node[anchor=south]{$z_2$};
        \draw (11.35, 4.85) circle (0cm) node[anchor=south]{$r_2$};

        \draw (9, 6.1) circle (0cm) node[anchor=south]{$x_3$};
        \draw (10.5, 6.1) circle (0cm) node[anchor=south]{$y_3$};
        \draw (12, 6.1) circle (0cm) node[anchor=south]{$z_3$};
        \draw (11.35, 6.35) circle (0cm) node[anchor=south]{$r_3$};

        \draw (9, 8.1) circle (0cm) node[anchor=south]{$x_{D}$};
        \draw (10.5, 8.1) circle (0cm) node[anchor=south]{$y_{D}$};
        \draw (12, 8.1) circle (0cm) node[anchor=south]{$z_{D}$};
        \draw (11.35, 8.35) circle (0cm) node[anchor=south]{$r_D$};

        \draw (13.65, 7.75) circle (0cm) node[anchor=south]{$Q_{D}$};
        \draw (13.65, 5.75) circle (0cm) node[anchor=south]{$Q_3$};
        \draw (13.65, 4.25) circle (0cm) node[anchor=south]{$Q_2$};
        \draw (13.65, 2.75) circle (0cm) node[anchor=south]{$Q_1$};

        \draw (4.5, 3.5) circle (0cm) node[anchor=south]{$G$};
        \draw (8, 0.25) circle (0cm) node[anchor=south]{$G'$};
    \end{tikzpicture}
    \caption{Construction of $G'$ from $G$. Here, $D$ is used as a placeholder for $d(v)+1$.}
    \label{fig: Figsolsize}
\end{figure}

The complexity result related to DS problem from the literature is given as follows. 
\begin{theorem} [\cite{raman}] ~\label{bp}
     DS problem on bipartite graphs is W[2]-hard parameterized by solution size.
\end{theorem}
\noindent \textbf{Construction. }Given an instance $I = (G, k)$ of DS problem, we transform it into an instance $I' = (G', k)$ of SRD problem as follows. For each vertex $v \in V(G)$, we construct a gadget as follows. We add exactly $d(v) + 1$ paths of length two. The paths are denoted by $P_1, P_2, ...,$ and $P_{d(v)+1}$. The vertices of path $P_i$ are denoted by $x_i, y_i$ and $z_i$ with $x_i$ adjacent to $v$. We add two pendant vertices adjacent to $z_i$ for $i \in \{1, 2, ..., d(v)+1\}$. The set containing the pendant vertices adjacent to $z_i$ is denoted by $Q_i$. We add two pendant vertices adjacent to $y_1$ and one pendant vertex adjacent to the vertices $y_i$ for $i \in \{2, ..., d(v)+1\}$. The set containing the pendant vertices adjacent to $y_1$ is denoted by $R_1$ and the pendant vertex adjacent to $y_i$ for $i \in \{2, ..., d(v)+1\}$ is denoted by $r_i$. This concludes the construction of $I'$. See Figure \ref{fig: Figsolsize} for an illustration. 

\begin{lemma}~\label{secondproof}
    $I$ has a dominating set of size at most $k$ if and only if $I'$ has signed Roman dominating function of weight at most $k$.
\end{lemma} 
\begin{proof}
        ($\Rightarrow$) Let $S \subseteq V$ be a dominating set of size at most $k$ and $f$ be a weighted signed Roman dominating function. Even though there exists a gadget for each vertex in $V(G)$, for simpler notations, we consider only one gadget that corresponds to a vertex $u \in V(G)$ and argue regarding the labels that the vertices of the gadget could take in $f$. The labels remain the same for each such gadget. We consider a vertex $u \in V(G)$ and its corresponding gadget in this proof. We use $D$ in place of $d(u)+1$.
        
        The corresponding signed Roman dominating function $f$ is given as follows.
\[
f(w) = 
\begin{cases}
-1, & \text{if } w \in \left( \bigcup\limits_{i \in \{1,2,...,D\}} \{x_i\}  \cup  \bigcup\limits_{i \in \{1,2,...,D\}} Q_i  \cup R_1 \cup \bigcup\limits_{i \in \{2,3,...,D\}} \{r_i\} \right), \\
1, & \text{if } w \in \left( v \mid v \notin S \right), \\
2, & \text{if } w \in \left(S \cup  \bigcup\limits_{i \in \{1,2,...,D\}} \{y_i, z_i\} \right)
\end{cases}
\]
    \begin{itemize}
        \item Each vertex from $\bigcup\limits_{i \in \{1,2,...,D\}} x_i$ has a positive labelSum due to the positive labels of both its neighbours $y_i$ and $u$. As the vertices $\bigcup\limits_{i \in \{1,2,...,D\}} x_i$ have the label $-1$ in $f$, they have a neighbour $y_i$ with label two.
        \item Each vertex from $\bigcup\limits_{i \in \{1,2,...,D\}} y_i$ has a positive labelSum due the label two of itself and its neighbour $z_i$. 
        \item Each vertex from $\bigcup\limits_{i \in \{1,2,...,D\}} z_i$ has a positive labelSum due to the label two of itself and its neighbor $y_i$.
        \item Each vertex from $
        \bigcup\limits_{i \in \{1,2,...,D\}} Q_i$ has a positive labelSum due to the label two of its neighbour $z_i$. As the vertices of $
        \bigcup\limits_{i \in \{1,2,...,D\}} Q_i$ have the label two in $f$, their only neighbour $z_i$ has the label two. 
        \item Each vertex from $R_1 \cup \bigcup\limits_{i \in \{2,...,D\}} r_i$ has a positive labelSum due to the label two of its only neighbor $y_i$. As the vertices of $
        R_1 \cup \bigcup\limits_{i \in \{2,...,D\}} r_i$ have the label $-1$ in $f$, their only neighbour $y_i$ has the label two. 
        \item Vertex $u \in V(G)$ has a labelsum of $-|D|-1$ from the gadget, and a labelSum of at least $|D|+2$ from its closed neighbours $N_G[u]$. Hence, $u$ has a positive labelSum.
    \end{itemize}
    Each vertex $w \in V(G')$ has a positive labelSum and each vertex $w \in V(G')$ with $f(w)=-1$ has a vertex $v \in N(w)$ with $f(v)=2$. Weight of the gadget adjacent to $u \in V(G)$ is $-1$. There are $n$ such gadgets (one for each vertex in $V(G)$), with a total weight of $-n$. At most $k$ vertices from $V(G)$ get the label two while the rest $n-k$ get the label one. The weight of $V(G)$ becomes $n+k$. Hence, the weight of the signed Roman dominating function $f$ will be at most $k$. 
    \medskip

    \noindent
($\Leftarrow$) Let $f$ be the minimum weighted signed Roman dominating function of weight at most $k$.
        \begin{itemize}
        \item Each vertex $y_i$ and $z_i$ must be assigned the label two in $f$ as the pendant vertices adjacent to these vertices are assigned the label $-1$. The weight of the set $\{y_i, z_i, r_i\} \cup Q_i$ for $i \in \{2, 3, ..., D\}$ is one. If we chose to label any pendant vertex from $\{r_i\} \cup Q_i$ with 1 or 2, then we need a weight of at least two for the set $\{y_i, z_i, r_i\} \cup Q_i$. The weight of the set $\{y_1, z_1\} \cup R_1 \cup Q_1$ is zero. If we chose to label any pendant vertex from $R_1 \cup Q_1$ with 1 or 2, then we need a weight of at least two for the set $\{y_1, z_1, r_1\} \cup Q_1$. Hence, in $f$, each pendant vertex must be assigned the label $-1$.
        \item At this point, the total weight of the set $\{y_1, z_1\} \cup R_1 \cup Q_1 \cup \bigcup\limits_{i \in \{2,3,...,D\}}\{y_i, z_i, r_i\} \cup Q_i$ is $D-1$. In order, to bring down the weight of each gadget, we must label the vertices in $\bigcup\limits_{i \in \{1,2,...,D\}} x_i$ with $-1$. The reason why any vertex from $\bigcup\limits_{i \in \{1,2,...,D\}} x_i$ do not get a positive label, will be explained later in this proof.
        \item After assigning the label $-1$ to each vertex in $\bigcup\limits_{i \in \{1,2,...,D\}} x_i$, the total weight of the gadget becomes $-1$. 
        \item Consider a vertex $x_i$ adjacent to the vertex $u \in V(G)$. As $x_i$ has a labelSum of one from its closed neighbours in the gadget, vertex $u$ cannot have the label $-1$. Hence, each vertex $u \in V(G)$ is assigned one among the labels from $\{1, 2\}$. 
        \item As the weight of $f$ is at most $k$ and the weight of the gadget adjacent to each vertex is $-1$, the weight of $V(G)$ will be at most $n+k$ in $G'$.
        \item At most $k$ vertices in $V(G)$ are assigned the label 2, while the rest are assigned the label 1.
        \item At this point, each vertex $u \in V(G)$ with a degree $D$ has a labelSum of $-(D+1)$ from its neighbours in the gadget. In order for $u$ to acquire a positive labelSum, at least one of its closed neighbours from $V(G)$ must be assigned the label 2. As we label at most $k$ vertices from $V(G)$ with 2, the overall weight of $f$ will be at most $k$.
        \item Let us assume that there exists some vertex from $\bigcup\limits_{i \in \{1,2,...,D\}} x_i$ with the label 1, then the weight of $V(G)$ should not be more than $n+k-2$. We can label at most $k-2$ vertices from $V(G)$ with 2. This leads to at least one vertex other than $u$ from $V(G)$ to have non-positive labelSum. Hence, we conclude that each vertex in $\bigcup\limits_{i \in \{1,2,...,D\}} x_i$ must be assigned the label $-1$ to obtain a minimum weighted signed Roman dominating function $f$.
        \item Finally, we conclude that the vertices of $G$ in $G'$ that receive the label 2 in $f$ will correspond to the dominating set in $G$.
    \end{itemize}
\end{proof}
\begin{claim} ~\label{bpclaim}
    If $G$ is a bipartite graph then $G'$ is a bipartite graph.
\end{claim}
\begin{proof}
        Let $G = A \cup B$ and $u \in A$. The corresponding bipartition after adding the gadget to vertex $u$ is $\{A \cup \bigcup\limits_{i \in \{1,2,..., D\}} y_i \cup P_Z\}$ $\cup$ $\{B \cup \bigcup\limits_{i \in \{1,2,..., D\}} (x_i \cup z_i) \cup P_Y\}$, where $P_Y$ represents the pendant vertices of $Y$ and $P_Z$ represents the pendant vertices of $Z$. Similarly, we can form the bipartition after adding the gadget to all the vertices of $G$. Hence, $G'$ is bipartite. 
\end{proof}
From Lemma \ref{secondproof}, Claim \ref{bpclaim} and Theorem \ref{bp}, we have the following result.
\begin{theorem}
    SRD problem on bipartite graphs is W[2]-hard parameterized by weight.
\end{theorem}
The result related to DS problem on circle graphs is given as follows.
\begin{theorem} [\cite{nicolas}] ~\label{circle}
    DS problem on circle graphs is W[1]-hard parameterized by solution size. 
\end{theorem} 
\begin{claim}~\label{circleclaim}
        If $G$ is a circle graph then $G'$ is a circle graph.
\end{claim}
\begin{proof}
        The circle representation of $G'$ with the gadget added for vertex $u$ is represented in Figure \ref{fig: Figcircle}. As the pendant vertices can be drawn without any crossings, they are not shown in the figure. In a similar way, we can represent the gadget for each vertex of $V(G)$, in a circle model.
        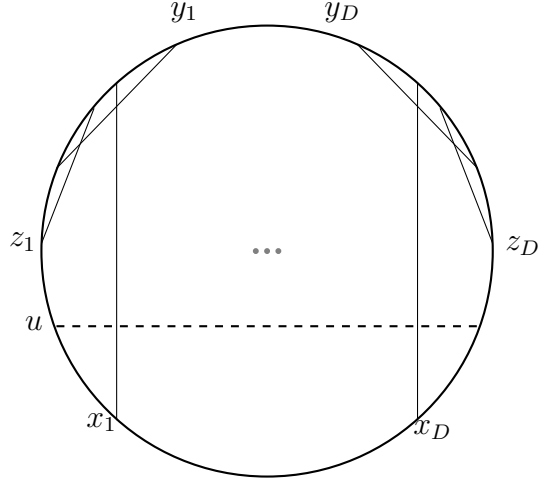
\begin{figure} [t]
\centering
    \begin{tikzpicture} [thick,scale=1, every node/.style={scale=1}]
        \draw (5, 5) circle (3);

        \draw[thick, dashed] (2.2, 4) -- (7.8, 4);

        \draw[thin] (3, 7.25) -- (3, 2.75); 
        
        \draw[thin] (2.2, 6.1) -- (3.8, 7.75);
        \draw[thin] (2, 5.1) -- (2.7, 6.9);

        \draw[thin] (7, 7.25) -- (7, 2.75); 
        
        \draw[thin] (7.8, 6.1) -- (6.2, 7.75);
        \draw[thin] (8, 5.1) -- (7.3, 6.9);

        \filldraw[gray] (4.85, 5) circle (0.025);        
        \filldraw[gray] (5, 5) circle (0.025);        
        \filldraw[gray] (5.15, 5) circle (0.025);
        
        \draw (1.9, 3.8) circle (0cm) node[anchor=south]{$u$};

        \draw (2.8, 2.45) circle (0cm) node[anchor=south]{$x_1$};
        \draw (3.9, 7.9) circle (0cm) node[anchor=south]{$y_1$};
        \draw (1.75, 4.85) circle (0cm) node[anchor=south]{$z_1$};

        \draw (7.2, 2.35) circle (0cm) node[anchor=south]{$x_D$};
        \draw (6, 7.9) circle (0cm) node[anchor=south]{$y_D$};
        \draw (8.4, 4.8) circle (0cm) node[anchor=south]{$z_D$};
    \end{tikzpicture}
    \caption{Circle representation of $G'$}
    \label{fig: Figcircle}
\end{figure}
\end{proof}
\noindent Similarly, from Lemma \ref{secondproof}, Claim \ref{circleclaim} and Theorem \ref{circle}, we obtain the following theorem.
\begin{theorem}
     SRD problem on circle graphs is W[1]-hard parameterized by weight. 
\end{theorem}

\section{W[1]-hard Parameterized by Feedback Vertex Set Number}
In this section, we study the parameterized complexity of SRD problem parameterized by feedback vertex set number. 

 We provide a parameterized reduction from \mrss{} (MRSS) problem to prove that SRD problem is W[1]-hard. The MRSS problem is defined as follows.
\begin{tcolorbox}
{
\mrss{}: \newline
\textit{Input:} An integer $k$, a set $S = \{s_1, s_2, ..., s_n\}$ of vectors with $s_i \in \mathbb{N}^k$ for every $i$ with $1 \leq i \leq n$, a target vector $t \in \mathbb{N}^k$ and an integer $m$. \newline
\textit{Parameter:} $k+m$\newline
\textit{Output:} Does there exist a subset $S' \subseteq S$ with $|S'| \leq m$ such that $\sum\limits_{s \in S'} s \geq t$?
}
\end{tcolorbox}
MRSS problem is known to be W[1]-hard parameterized by $k+m$, even when all integers are given in unary~\cite{GNN}.
\medskip

\noindent \textbf{Construction.} Let $I = (k, m, S, t)$ be an instance of MRSS problem and for $s_i \in S$, $max(s_i)$ denotes the value of the largest coordinate of $s_i$. We construct the reduced instance $I' = (G, k')$ of SRD problem as follows. 
\begin{itemize}
    \item We create two sets of $k$ vertices $\{u_1,u_2,...,u_k\}$ and $\{v_1,v_2,...,v_k\}$. 
    \item We introduce a set $D_j$ with $(\sum\limits_{s_i \in S} s_i(j))+t(j)$ vertices, for each $j \in \{1,2,...,k\}$.  
    \item We introduce a set $F_j$ with $\lceil \frac{(\sum\limits_{s_i \in S} s_i(j))+t(j)}{2} \rceil$ vertices, for each $j \in \{1,2,...,k\}$. We make two pendant vertices adjacent to each vertex of $F_j$. The set containing the pendant vertices of $F_j$ is denoted by $P_j$. 
    \item For each $j \in \{1,2,...,k\}$, we make one pendant vertex adjacent to $u_j$ and $v_j$. The pendant vertex adjacent to $u_j$ and $v_j$ are denoted by $r_j^1$ and $r_j^2$, respectively. 
    \item For each $j \in \{1,2,...,k\}$, we make $v_j$ adjacent to each vertex in the sets $D_j$ and $F_j$. We also make $u_j$ adjacent to each vertex in the set $D_j$. 
    \item For each $s_i \in S$, we create $max(s_i)$ paths of length one. The vertices of each path is denoted by $b_i^l$ and $c_i^l$ where $l \in \{1, 2, ..., max(s_i)\}$.  
    \item Let $B_{s_i} = \bigcup\limits_{l \in \{1,2, ..., max(s_i)\}} b_i^l$ and $C_{s_i} = \bigcup\limits_{l \in \{1,2, ..., max(s_i)\}} c_i^l$. 
    \item We also create a set of vertices $\{a_1, a_2, ..., a_n\}$. Vertex $a_i$ will be made adjacent to each vertex in the set $B_{s_i}$. 
    \item For each $s_i \in S$ and $j \in \{1,2, ..., k\}$, we make $u_j$ adjacent to exactly $s_i(j)$ vertices in $C_{s_i}$ arbitrarily. 
    \item We make four pendant vertices adjacent to each vertex $b_i^l \in B_{s_i}$, all of which are denoted by $Z_i^l$.  
    \item We also create a path of length two containing the vertices $w_i^l, x_i^l$ and $y_i^l$ for $i \in \{1,2,...,n\}$ and $l \in \{1,2,...,max(s_i)\}$. We make $w_i^l$ adjacent to $b_i^l$ and $x_i^l$ and we make $x_i^l$ adjacent to $y_i^l$.  
    \item Let $W_{s_i}$ $ = \bigcup\limits_{l \in \{1,2, ..., max(s_i)\}} $ $w_i^l$, $X_{s_i}$ $ = \bigcup\limits_{l \in \{1,2, ..., max(s_i)\}} $ $x_i^l$, $Y_{s_i}$ $ = \bigcup\limits_{l \in \{1,2, ..., max(s_i)\}} $ $y_i^l$ and $Z_{s_i}$ $ = \bigcup\limits_{l \in \{1,2, ..., max(s_i)\}} $ $Z_i^l$.  
    \item For each vertex $c_i^l \in C_{s_i}$, we create two paths of two vertices each. The vertices of the first path are denoted by $g_i^l$ and $h_i^l$ and the vertices of the second path are denoted by $p_i^l$ and $q_i^l$. We make $c_i^l$ adjacent to $g_i^l$ and $p_i^l$.
    \item Let $G_{s_i}$ $ = \bigcup\limits_{l \in \{1,2, ..., max(s_i)\}}$ $ g_i^l$, $H_{s_i}$ $ = \bigcup\limits_{l \in \{1,2, ..., max(s_i)\}}$ $ h_i^l$, $P_{s_i}$ $ = \bigcup\limits_{l \in \{1,2, ..., max(s_i)\}}$ $ p_i^l$ and $Q_{s_i}$ $ = \bigcup\limits_{l \in \{1,2, ..., max(s_i)\}}$ $q_i^l$. 
\end{itemize}
\begin{figure} [t]
\centering
    \begin{tikzpicture} [thick,scale=0.35, every node/.style={scale=0.8}]
    
        \filldraw (2, -1) circle (0.15cm);
        \filldraw (8, -1) circle (0.15cm);

        \filldraw (16, -1) circle (0.15cm);
        \filldraw (22, -1) circle (0.15cm);

        \filldraw (31, -1) circle (0.15cm);

        \filldraw (5, -4) circle (0.15cm);
        \filldraw (19, -4) circle (0.15cm);
        \filldraw (31, -4) circle (0.15cm);
        
        \draw[thin] (2, 1.9) -- (2, -0.9); 
        \draw[thin] (8, 1.9) -- (8, -0.9); 

        \draw[thin] (16, 1.9) -- (16, -0.9); 
        \draw[thin] (22, 1.9) -- (22, -0.9); 

        \draw[thin] (31, 1.9) -- (31, -0.9); 

        \draw[thin] (2, -1.1) -- (5, -3.9); 
        \draw[thin] (8, -1.1) -- (5, -3.9); 

        \draw[thin] (16, -1.1) -- (19, -3.9); 
        \draw[thin] (22, -1.1) -- (19, -3.9); 

        \draw[thin] (31, -1.1) -- (31, -3.9); 
        
        \filldraw (2, 2) circle (0.15cm);
        \filldraw (8, 2) circle (0.15cm);
        \filldraw (16, 2) circle (0.15cm);
        \filldraw (22, 2) circle (0.15cm);
        \filldraw (31, 2) circle (0.15cm);

        \filldraw (5, 9) circle (0.15cm);
        \filldraw (31, 9) circle (0.15cm);

        \draw[thin] (2.07, 2.07) -- (4.93, 8.93); 
        \draw[thin] (7.93, 2.07) -- (5.07, 8.93); 
        \draw[thin] (15.93, 2.07) -- (5.07, 8.93);
        \draw[thin] (30.93, 2.07) -- (5.07, 8.93);

        \draw[thin] (2.07, 2.07) -- (30.93, 8.93); 
        \draw[thin] (16.07, 2.07) -- (30.93, 8.93);
        \draw[thin] (22.07, 2.07) -- (30.93, 8.93);
        \draw[thin] (31, 2.1) -- (31, 8.9);
        
        \filldraw (5, 12) circle (0.15cm);
        \filldraw (3.5, 12) circle (0.15cm);
        \filldraw (6.5, 12) circle (0.15cm);
        \filldraw (2, 12) circle (0.15cm);
        \filldraw (8, 12) circle (0.15cm);
        \filldraw (0.5, 12) circle (0.15cm);
        \filldraw (9.5, 12) circle (0.15cm);

        \draw[thin] (5, 11.9) -- (5, 9.1);
        \draw[thin] (3.57, 11.93) -- (4.93, 9.07);
        \draw[thin] (6.43, 11.93) -- (5.07, 9.07);
        \draw[thin] (2.07, 11.93) -- (4.93, 9.07);
        \draw[thin] (7.93, 11.93) -- (5.07, 9.07);
        \draw[thin] (0.57, 11.93) -- (4.93, 9.07);
        \draw[thin] (9.43, 11.93) -- (5.07, 9.07);
        
        \filldraw (31, 12) circle (0.15cm);
        \filldraw (29.5, 12) circle (0.15cm);
        \filldraw (32.5, 12) circle (0.15cm);
        \filldraw (28, 12) circle (0.15cm);
        \filldraw (34, 12) circle (0.15cm);
        \filldraw (26.5, 12) circle (0.15cm);
        \filldraw (35.5, 12) circle (0.15cm);

        \draw[thin] (31, 11.9) -- (31, 9.1);
        \draw[thin] (29.57, 11.93) -- (30.93, 9.07);
        \draw[thin] (32.43, 11.93) -- (31.07, 9.07);
        \draw[thin] (28.07, 11.93) -- (30.93, 9.07);
        \draw[thin] (33.93, 11.93) -- (31.07, 9.07);
        \draw[thin] (26.57, 11.93) -- (30.93, 9.07);
        \draw[thin] (35.43, 11.93) -- (31.07, 9.07);
        
        \filldraw (5, 15) circle (0.15cm);
        \filldraw (31, 15) circle (0.15cm);

        \filldraw (1, 18) circle (0.15cm);
        \filldraw (6.5, 18) circle (0.15cm);
        \filldraw (9, 18) circle (0.15cm);
        \filldraw (3.5, 18) circle (0.15cm);

        \draw[thin] (1.07, 17.93) -- (4.93, 15.07);
        \draw[thin] (8.93, 17.93) -- (5.07, 15.07);
        \draw[thin] (3.57, 17.93) -- (4.93, 15.07);
        \draw[thin] (6.43, 17.93) -- (5.07, 15.07);
        
        \filldraw (29.5, 18) circle (0.15cm);
        \filldraw (32.5, 18) circle (0.15cm);
        \filldraw (27, 18) circle (0.15cm);
        \filldraw (35, 18) circle (0.15cm);

        \draw[thin] (27.07, 17.93) -- (30.93, 15.07);
        \draw[thin] (32.43, 17.93) -- (31.07, 15.07);
        \draw[thin] (29.57, 17.93) -- (30.93, 15.07);
        \draw[thin] (34.93, 17.93) -- (31.07, 15.07);

        \filldraw (0.5, 20) circle (0.15cm);
        \filldraw (1.5, 20) circle (0.15cm);
        \filldraw (3, 20) circle (0.15cm);
        \filldraw (4, 20) circle (0.15cm);
        \filldraw (6, 20) circle (0.15cm);
        \filldraw (7, 20) circle (0.15cm);
        \filldraw (8.5, 20) circle (0.15cm);
        \filldraw (9.5, 20) circle (0.15cm);

        \draw[thin] (0.57, 19.93) -- (0.93, 18.07);
        \draw[thin] (1.43, 19.93) -- (1.07, 18.07);

        \draw[thin] (3.07, 19.93) -- (3.43, 18.07);
        \draw[thin] (3.93, 19.93) -- (3.57, 18.07);


        \draw[thin] (6.07, 19.93) -- (6.43, 18.07);
        \draw[thin] (6.93, 19.93) -- (6.57, 18.07);

        \draw[thin] (8.57, 19.93) -- (8.93, 18.07);
        \draw[thin] (9.43, 19.93) -- (9.07, 18.07);

        \filldraw (26.5, 20) circle (0.15cm);
        \filldraw (27.5, 20) circle (0.15cm);
        \filldraw (29, 20) circle (0.15cm);
        \filldraw (30, 20) circle (0.15cm);
        \filldraw (32, 20) circle (0.15cm);
        \filldraw (33, 20) circle (0.15cm);
        \filldraw (34.5, 20) circle (0.15cm);
        \filldraw (35.5, 20) circle (0.15cm);

        \draw[thin] (26.57, 19.93) -- (26.93, 18.07);
        \draw[thin] (27.43, 19.93) -- (27.07, 18.07);

        \draw[thin] (29.07, 19.93) -- (29.43, 18.07);
        \draw[thin] (29.93, 19.93) -- (29.57, 18.07);


        \draw[thin] (32.07, 19.93) -- (32.43, 18.07);
        \draw[thin] (32.93, 19.93) -- (32.57, 18.07);

        \draw[thin] (34.57, 19.93) -- (34.93, 18.07);
        \draw[thin] (35.43, 19.93) -- (35.07, 18.07);

        \draw[thin] (5, 14.9) -- (5, 12.1);
        \draw[thin] (4.93, 14.93) -- (3.5, 12.1);
        \draw[thin] (5.07, 14.93) -- (6.5, 12.1);
        \draw[thin] (4.93, 14.93) -- (2, 12.1);
        \draw[thin] (5.07, 14.93) -- (8, 12.1);
        \draw[thin] (4.93, 14.93) -- (0.5, 12.1);
        \draw[thin] (5.07, 14.93) -- (9.5, 12.1);
        
        \draw[thin] (31, 14.9) -- (31, 12.1);
        \draw[thin] (30.93, 14.93) -- (29.5, 12.1);
        \draw[thin] (31.07, 14.93) -- (32.5, 12.1);
        \draw[thin] (30.93, 14.93) -- (28, 12.1);
        \draw[thin] (31.07, 14.93) -- (34, 12.1);
        \draw[thin] (30.93, 14.93) -- (26.5, 12.1);
        \draw[thin] (31.07, 14.93) -- (35.5, 12.1);

        \draw (1, 1.5) circle (0cm) node[anchor=south]{$c_1^1$};
        \draw (15, 1.5) circle (0cm) node[anchor=south]{$c_2^1$};
        \draw (32, 1.5) circle (0cm) node[anchor=south]{$c_3^1$};


        \draw (9, 1.5) circle (0cm) node[anchor=south]{$c_1^2$};
        \draw (23, 1.5) circle (0cm) node[anchor=south]{$c_2^2$};

        \draw (1, -1.5) circle (0cm) node[anchor=south]{$b_1^1$};
        \draw (15, -1.5) circle (0cm) node[anchor=south]{$b_2^1$};
        \draw (32, -1.5) circle (0cm) node[anchor=south]{$b_3^1$};


        \draw (9, -1.5) circle (0cm) node[anchor=south]{$b_1^2$};
        \draw (23, -1.5) circle (0cm) node[anchor=south]{$b_2^2$};

        \draw (5, -5.25) circle (0cm) node[anchor=south]{$a_1$};
        \draw (19, -5.25) circle (0cm) node[anchor=south]{$a_2$};
        \draw (31.5, -5.25) circle (0cm) node[anchor=south]{$a_3$};

        \draw (4, 8.5) circle (0cm) node[anchor=south]{$u_1$};
        \draw (30, 8.5) circle (0cm) node[anchor=south]{$u_2$};

        \draw (-0.5, 11.5) circle (0cm) node[anchor=south]{$D_1$};
        \draw (25.5, 11.5) circle (0cm) node[anchor=south]{$D_2$};

        \draw (4, 14.5) circle (0cm) node[anchor=south]{$v_1$};
        \draw (30, 14.5) circle (0cm) node[anchor=south]{$v_2$};

        \draw (0, 17.5) circle (0cm) node[anchor=south]{$F_1$};
        \draw (26, 17.5) circle (0cm) node[anchor=south]{$F_2$};
        
        \draw (-0.5, 19.5) circle (0cm) node[anchor=south]{$P_1$};
        \draw (25.5, 19.5) circle (0cm) node[anchor=south]{$P_2$};
        

        \filldraw (7, 15) circle (0.15cm);
        \filldraw (33, 15) circle (0.15cm);

        \draw[thin] (5.1, 15) -- (6.9, 15);
        \draw[thin] (31.1, 15) -- (32.9, 15);

        \draw (8, 14.5) circle (0cm) node[anchor=south]{$r_1^2$};
        \draw (34, 14.5) circle (0cm) node[anchor=south]{$r_2^2$};

        \filldraw (7, 9) circle (0.15cm);
        \filldraw (33, 9) circle (0.15cm);

        \draw[thin] (5.1, 9) -- (6.9, 9);
        \draw[thin] (31.1, 9) -- (32.9, 9);

        \draw (8, 8.5) circle (0cm) node[anchor=south]{$r_1^1$};
        \draw (34, 8.5) circle (0cm) node[anchor=south]{$r_2^1$};
        
        \draw (19, -8.75) circle (0cm) node[anchor=south]{$G$};

    \end{tikzpicture}
    \caption{Reduced instance of SRD problem constructed from MRSS problem instance $S =\{(2, 1), (1, 2), (1, 1)\}, t= (3, 3), k = 2, m = 2.$}
    \label{fig: Figfvs1}
\end{figure}
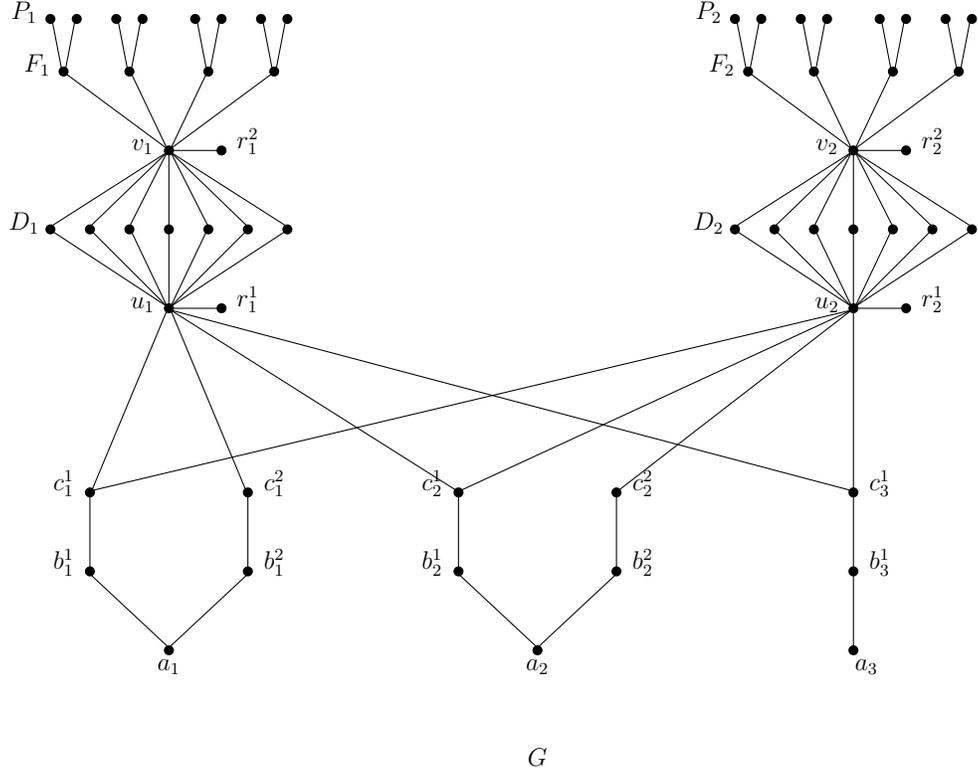
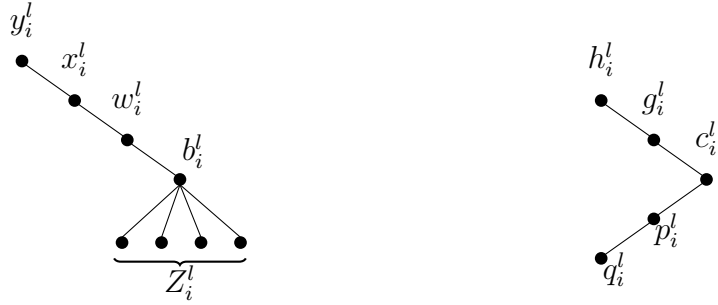
\begin{figure} [t]
\centering
    \begin{tikzpicture} [thick,scale=0.7, every node/.style={scale=1}]

        \draw (18, 3.5) circle (0cm) node[anchor=south]{};
    
        \filldraw (22, 0) circle (0.1cm);
        
        \filldraw (21, 0.75) circle (0.1cm);
        \filldraw (20, 1.5) circle (0.1cm);
        
        \filldraw (21, -0.75) circle (0.1cm);
        \filldraw (20, -1.5) circle (0.1cm);

        \draw[thin] (21.93, 0.07) -- (21.07, 0.68);
        \draw[thin] (21.93, -0.07) -- (21.07, -0.68);

        \draw[thin] (20.93, 0.82) -- (20.07, 1.43);
        \draw[thin] (20.93, -0.82) -- (20.07, -1.43);

        \draw (22, 0.25) circle (0cm) node[anchor=south]{$c_i^l$};
        \draw (21, 1) circle (0cm) node[anchor=south]{$g_i^l$};
        \draw (20, 1.75) circle (0cm) node[anchor=south]{$h_i^l$};
        \draw (21.25, -1.5) circle (0cm) node[anchor=south]{$p_i^l$};
        \draw (20.25, -2.25) circle (0cm) node[anchor=south]{$q_i^l$};

        \filldraw (12, 0) circle (0.1cm);
        
        \filldraw (11, 0.75) circle (0.1cm);
        \filldraw (10, 1.5) circle (0.1cm);
        \filldraw (9, 2.25) circle (0.1cm);
        
        \filldraw (10.9, -1.2) circle (0.1cm);
        \filldraw (11.65, -1.2) circle (0.1cm);
        \filldraw (12.4, -1.2) circle (0.1cm);
        \filldraw (13.15, -1.2) circle (0.1cm);

        \draw[thin] (11.93, 0.07) -- (11.07, 0.68);
        \draw[thin] (10.07, 1.43) -- (10.93, 0.82);
        \draw[thin] (9.93, 1.57) -- (9.07, 2.18);

        \draw[thin] (12, -0.1) -- (10.9, -1.1);
        \draw[thin] (12, -0.1) -- (11.65, -1.1);
        \draw[thin] (12, -0.1) -- (12.4, -1.1);
        \draw[thin] (12, -0.1) -- (13.15, -1.1);

        \draw (12.25, 0) circle (0cm) node[anchor=south]{$b_i^l$};
        \draw (11, 1) circle (0cm) node[anchor=south]{$w_i^l$};
        \draw (10, 1.75) circle (0cm) node[anchor=south]{$x_i^l$};
        \draw (9, 2.5) circle (0cm) node[anchor=south]{$y_i^l$};

        \draw (12, -2.5) circle (0cm) node[anchor=south]{$Z_i^l$};
        
        \draw [decorate, decoration = {brace}]  (13.25, -1.5) -- (10.75, -1.5);
    \end{tikzpicture}
    \caption{Gadgets for the vertices of the sets $B_{s_i}$ (on the left) and $C_{s_i}$ (on the right).}
    \label{fig: Figfvs2}
\end{figure} 

This completes the construction of the SRD problem instance $I'$. For more details, see Figure \ref{fig: Figfvs1} and Figure \ref{fig: Figfvs2}.

We set $k' = \sum\limits_{s_i \in S}$ ($3 \cdot max(s_i)$+1) $-\sum\limits_{j \in \{1,2,...,k\}}((\sum\limits_{s_i \in S} s_i(j))+t(j))$ $+2k+m$ and the feedback vertex set is $\bigcup\limits_{j \in \{1,2, ..., k\}} (u_j \cup v_j)$ of size $2k$.
\begin{lemma}~\label{forward}
    \textit{If ($k, m, S, t$) is a yes instance of }MRSS problem\textit{ then there exists a signed Roman dominating function of weight at most $k'$.} 
\end{lemma}
\begin{proof}
        Let $S' \subseteq S$ such that $|S'| \leq m$ and $\sum\limits_{s_i \in S'}s_i \geq t$. 
        
        The corresponding signed Roman dominating function $f$ is given as follows. 
        
        \[f(w) =\begin{cases}
            -1, & \text{if } w \in (\bigcup\limits_{j \in \{1,2, ..., k\}} (P_j \cup D_j \cup \{r_j^1,r_j^2\} \cup \bigcup\limits_{i \in \{1,2,...,n\}} (H_{s_i} \cup Q_{s_i} \cup Z_{s_i}) \\ &\cup \bigcup\limits_{\substack{i \in \{1,2,...,n\}\\ s_i \in S'}} W_{s_i} \cup \bigcup\limits_{\substack{i \in \{1,2,...,n\}\\s_i \notin S'}} Y_{s_i}), \\
    1, & \text{if } w \in \left(\bigcup\limits_{\substack{i \in \{1,2,...,n\}\\ s_i \in S'}} (X_{s_i} \cup Y_{s_i}) \cup \bigcup\limits_{\substack{i \in \{1,2,...,n\} \\ s_i \notin S'}} (W_{s_i} \cup C_{s_i}  \cup \{a_i\})\right), \\
    2, & \text{if } w \in \left(\bigcup\limits_{j \in \{1,2, ..., k\}} (F_j \cup \{u_j, v_j\}) \cup \bigcup\limits_{i \in \{1,2,...,n\}} (B_{s_i} \cup G_{s_i} \cup P_{s_i})\right) \\ & \cup \left(\bigcup\limits_{\substack{i \in \{1,2,...,n\}\\ s_i \in S'}} (C_{s_i} \cup \{a_i\})\right) \cup \left(\bigcup\limits_{\substack{i \in \{1,2,...,n\} \\ s_i \notin S'}} X_{s_i}\right)
        \end{cases}\]
    \begin{itemize}
        \item As the vertices of $F_j$ have the label 2 and the vertices of $P_j$ have the label $-1$, each vertex of $P_j$ has a labelSum of exactly one and also has a neighbour with label 2. 
        \item Each vertex of $F_j$ has a labelSum of two due to the positive labels of the vertex itself and its neighbour $v_j$. 
        \item Vertex $v_j$ has a labelSum of at least one, the positive weight of $F_j$ is at least the negative weight of $D_j$ and the positive weight of one obtained from the combined vertices $v_j$ and $r_j^2$ will contribute to the positive labelSum of $v_j$. 
        \item Vertex $u_j$ is adjacent to $\sum\limits_{s_i \in S} s_i(j)$ vertices from $C$. Among this, at least $t(j)$ vertices gets the label 2 and at most $(\sum\limits_{s_i \in S} s_i(j))-t(j)$ gets the label 1. The weight of the vertices adjacent to $u_j$ in $C$ is at least $((\sum\limits_{s_i \in S} s_i(j))+t(j))$. The weight of $D_j$ is $-((\sum\limits_{s_i \in S} s_i(j))+t(j))$. The combined weight of the vertices adjacent to $u_j$ in $D_j \cup C$ is non-negative. The positive weight of one obtained from the combined vertices $u_j$ and $r_j^1$ will contribute to the positive labelSum of $u_j$.  
        \item Vertices $r_j^1$ and $r_j^2$ have positive labelSum as their only neighbour ($u_j$ and $v_j$, respectively) has the label 2. 
        \item Each vertex of $C_{s_i}$ has a positive labelSum as all its neighbours from $\bigcup\limits_{j \in \{1,2,...,k\}} u_j$ have the label 2. 
        \item Each vertex of $B_{s_i}$ has a labelSum of eactly one, the labels of its neighbours $c_i^l, w_i^l$ and $a_i$ depends on whether $s_i \in S'$. 
        \item Vertex $a_i$ has a positive labelSum as all its neighbours from $B_{s_i}$ have the label 2. 
        \item Each vertex $w \in W_{s_i}$ has a positive labelSum as $b_i^l$ has the label 2, and one among $x_i^l$ and the vertex $w$ itself gets a positive label based on whether $s_i \in S'$. 
        \item Each vertex of $X_{s_i}$ has a positive labelSum as one of its neighbours $w_i^l$ and $y_i^l$ has a positive label based on whether $s_i \in S'$. 
        \item Each vertex of $Y_{s_i}$ has a labelSum of one as its only neighbour gets the label 1 or 2 based on whether $s_i \in S'$. 
        \item Each vertex of $Z_{s_i}$ has a labelSum of one, as their only neighbour $b_i^l$ has the label 2. 
        \item Each vertex of $G_{s_i}$ and $P_{s_i}$ has a positive labelSum due to the positive label of $c_i^l$ and the vertex itself. 
        \item Each vertex of $H_{s_i}$ and $Q_{s_i}$ has a positive labelSum due to the label 2 of its neighbour $g_i^l$ and $p_i^l$ respectively.
    \end{itemize}
Each vertex $w \in V(G')$ has a positive labelSum and each vertex $w \in V(G')$ with $f(w)=-1$ has a vertex $v \in N(w)$ with $f(v)=2$. Let $s_i \in S'$, the combined weight of the vertices $\{a_i\} \cup B_{s_i} \cup C_{s_i} \cup G_{s_i} \cup H_{s_i} \cup P_{s_i} \cup Q_{s_i} \cup W_{s_i} \cup X_{s_i} \cup Y_{s_i} \cup Z_{s_i}$ is $3 \cdot max(s_i)+2$. Let $s_i \notin S'$, the combined weight of the vertices $\{a_i\} \cup B_{s_i} \cup C_{s_i} \cup G_{s_i} \cup H_{s_i} \cup P_{s_i} \cup Q_{s_i} \cup W_{s_i} \cup X_{s_i} \cup Y_{s_i} \cup Z_{s_i}$ is $3 \cdot max(s_i)+1$. The weight of the set $F_j \cup P_j$ is zero. The weight of $D_j$ is $-((\sum\limits_{s_i \in S} s_i(j))+t(j))$. The weight of $\{u_j, v_j, r_j^1, r_j^2\}$ is two. Hence, the weight of the signed Roman dominating function $f$ is $k' = \sum\limits_{s_i \in S}$ ($3 \cdot max(s_i)+1$) $-\sum\limits_{j \in \{1,2,...,k\}}((\sum\limits_{s_i \in S} s_i(j))+t(j))$ $+2k+m$.
\end{proof}
\begin{lemma}~\label{obs1}
   Let $f$ a minimum weighted signed Roman dominating function. For each $j \in \{1,2, ..., k\}$, each vertex in $F_j$ is assigned the label 2 and each vertex in $P_j$ is assigned the label $-1$ in $f$.
\end{lemma} 
\begin{proof}
    Given a minimum weighted signed Roman dominating function $f$. There are three ways to label the vertex $v_j$ and its adjacent pendant vertex $r_j^2$ in $f$. 
\medskip

\noindent \textbf{Case 1.} $f(v_j) =2$ and $f(r_j^2)=-1$.

\noindent \textbf{Case 2.} $f(v_j) =$ $f(r_j^2) = 1$. 

\noindent  \textbf{Case 3.} $f(v_j) = -1$ and $f(r_j^2) = 2$.
\medskip

By swapping the labels of the vertices in case 3, we obtain case 1,  and it is easy to see that if there exist a minimum weighted $f$ with case 3 then there must exist a minimum weighted $f$ with case 1 as well. Hence, we conclude that there exists a minimum weighted $f$ from one among case 1 and case 2. As there exists a minimum weighted $f$ with $f(v_j) = 1$ or $f(v_j) = 2$, we label the vertices of $F_j$ with 2 and the pendant vertices from $P_j$ with $-1$. This leads to a minimum weight for $F_j \cup P_j$ and each vertex of $F_j$ has a positive labelSum. Hence, we conclude that there exists a minimum weighted $f$ with each vertex of $F_j$ with the label 2 and the vertices of $P_j$ with the label $-1$.
\end{proof}
\begin{lemma}~\label{obs2}
   Let $f$ a minimum weighted signed Roman dominating function. For each $j \in \{1,2, ..., k\}$, $v_j$ is assigned the label 2 and $r_j^2$ is assigned the label $-1$ in $f$. 
\end{lemma}
   \begin{proof}
       From Lemma \ref{obs1}, each vertex of $F_j$ has the label 2. The weight of $F_j \cup D_j$ is at least 0 (exactly zero if all the vertices of $D_j$ have the label $-1$). In order for $v_j$ to have a positive labelSum, the weight of $\{v_j, r_j^2\}$ must be at least $1$. We label the vertex $v_j$ with 2 and $r_j^2$ with $-1$, which is superior to labeling both $v_j$ and $r_j^2$ with 1. Hence, we conclude that there exists a minimum weighted $f$ with $v_j$ getting the label 2 and the adjacent pendant vertex $r_j^2$ getting the label $-1$.
   \end{proof}
\begin{lemma}~\label{weightcases}
Let $f$ be a minimum weighted signed Roman domination function. For each $s_i \in S$, the weight of $\{a_i\} \cup B_{s_i} \cup C_{s_i} \cup G_{s_i} \cup H_{s_i} \cup P_{s_i} \cup Q_{s_i} \cup W_{s_i} \cup X_{s_i} \cup Y_{s_i} \cup Z_{s_i}$ is 
\begin{enumerate}
    \item $4 \cdot max(s_i)+2$ if $f(v) = -1$ for each vertex $v \in C_{s_i}$.
    \item $3 \cdot max(s_i)+1$ if $f(v) = 1$ for each vertex $v \in C_{s_i}$.
    \item $3 \cdot max(s_i)+2$ if $f(v) = 2$ for each vertex $v \in C_{s_i}$.
    \item $3 \cdot max(s_i)+2$ if $f(v) = 2$ for some vertex $v \in C_{s_i}$ and $f(w) \neq -1$ for each vertex $w \in C_{s_i}$.
    \item $3 \cdot max(s_i)+5$ if $f(v) = -1$ for some vertex $v \in C_{s_i}$.
\end{enumerate}
\end{lemma}
\begin{proof}
    Let $f$ be a minimum weighted signed Roman dominating function. We fix the label of $c_i^l$ and then argue about the labels of other vertices in each case. 
    \medskip
    
\noindent \textbf{Case 1.} Let $f(c_i^l) = -1$ for each vertex $c_i^l \in C_{s_i}$.
\begin{itemize}
    \item The labeling for the vertices $f(g_i^l), f(h_i^l), f(p_i^l)$ and $ f(q_i^l)$ that will lead to a minimum weighted $f$ is $f(g_i^l) = 1, f(h_i^l) = 1, f(p_i^l) = 1, f(q_i^l) = 1$. This is true because, as $f(c_i^l) = -1$, for $f(g_i^l)$ to have a positive labelSum with minimum weight, we must set $f(g_i^l) =1$ and $f(h_i^l) = 1$. Same is the case for $p_i^l$ and $q_i^l$.
    \item As $b_i^l$ has four pendant vertices adjacent to it, to obtain a minimum weighted $f$, we set $f(b_i^l) = 2$ and $f(w) = -1$ for each vertex $w \in Z_i^l$. This leads to a minimum weighted $f$ because if we label $f(b_i^l) = -1$ and $f(z) = 2$ for each vertex $z \in Z_i^l$, weight of $f$ increases by at least four.
    \item As $b_i^l$ has a labelSum of $-3$ from its closed neighbours in $\{c_i^l, b_i^l\} \cup Z_i^l$, we set $f(w_i^l) = 2$ and $f(a_i) = 2$. 
    \item Accordingly, we get $f(x_i^l) = 2, f(y_i^l) = -1$.
\end{itemize}
The weight of $\{a_i\} \cup B_{s_i} \cup C_{s_i} \cup G_{s_i} \cup H_{s_i} \cup P_{s_i} \cup Q_{s_i} \cup W_{s_i} \cup X_{s_i} \cup Y_{s_i} \cup Z_{s_i}$, in this case is $4\cdot max(s_i)+2$. 
    \medskip
    
\noindent \textbf{Case 2.} Let $f(c_i^l) = 1$ for each vertex $c_i^l \in C_{s_i}$.
\begin{itemize}
    \item The labeling for the vertices $f(g_i^l), f(h_i^l), f(p_i^l)$ and $ f(q_i^l)$ that will lead to a minimum weighted $f$ is  $f(g_i^l) = 2, f(h_i^l) = -1, f(p_i^l) = 2, f(q_i^l) = -1$. As $f(c_i^l) = 1$, for $f(g_i^l)$ to have a positive labelSum either $f(g_i^l) =2$ and $f(h_i^l) = -1$ or $f(g_i^l) =1$ and $f(h_i^l) = 1$, the former labeling is superior to the latter due to the lesser weight of $\{g_i^l, h_i^l\}$. Same is the case for $p_i^l$ and $q_i^l$.
    \item Based on the argument in case 1 regarding the labels of $b_i^l$ and $Z_i^l$, we set $f(b_i^l) = 2$ and $f(z) = -1$ for each vertex $z \in Z_i^l$. 
    \item At this point, the labelSum of $b_i^l$ from its closed neighbours in $\{c_i^l, b_i^l\} \cup Z_i^l$ is $-1$, we set $f(w_i^l) = 1$ and $f(a_i) = 1$.  
    \item Similarly, we get $f(x_i^l) = 2, f(y_i^l) = -1$.
\end{itemize}
The weight of $\{a_i\} \cup B_{s_i} \cup C_{s_i} \cup G_{s_i} \cup H_{s_i} \cup P_{s_i} \cup Q_{s_i} \cup W_{s_i} \cup X_{s_i} \cup Y_{s_i} \cup Z_{s_i}$, in this case is $3\cdot max(s_i)+1$. 
    \medskip
    
\noindent \textbf{Case 3.} Let $f(c_i^l) = 2$ for each vertex $c_i^l \in C_{s_i}$.
\begin{itemize}
    \item The labeling for the vertices $f(g_i^l), f(h_i^l), f(p_i^l)$ and $ f(q_i^l)$ that will lead to a minimum weighted $f$ is  $f(g_i^l) = 2, f(h_i^l) = -1, f(p_i^l) = 2, f(q_i^l) = -1$. As $f(c_i^l) = 2$, for $f(g_i^l)$ to have a positive labelSum either $f(g_i^l) =2$ and $f(h_i^l) = -1$ or $f(g_i^l) =1$ and $f(h_i^l) = 1$, the former labeling is superior to the latter due to the lesser weight of $\{g_i^l, h_i^l\}$. Same is the case for $p_i^l$ and $q_i^l$.
    \item Based on the argument in case 1 regarding the labels of $b_i^l$ and $Z_i^l$, we set $f(b_i^l) = 2$ and $f(z) = -1$ for each vertex $z \in Z_i^l$. 
    \item At this point, the labelSum of $b_i^l$ from its closed neighbours in $\{c_i^l, b_i^l\} \cup Z_i^l$ is zero, we set $f(w_i^l) = -1$ and $f(a_i) = 2$. We can not set $f(a_i^l)$, as $a_i^l$ adjacent to each vertex of $b_i^l$. Setting both $f(w_i^l) = 1$ and $f(a_i) = 1$, would lead to a larger weight.
    \item As $f(w_i^l) = -1$, we get $f(x_i^l) = 1, f(y_i^l) = 1$.
\end{itemize}
The weight of $\{a_i\} \cup B_{s_i} \cup C_{s_i} \cup G_{s_i} \cup H_{s_i} \cup P_{s_i} \cup Q_{s_i} \cup W_{s_i} \cup X_{s_i} \cup Y_{s_i} \cup Z_{s_i}$, in this case is $3\cdot max(s_i)+2$. 
    \medskip
    
\noindent \textbf{Case 4.} Let $f(c_i^l) = 2$ for some vertex $c_i^l \in C_{s_i}$ and $f(c_i^l) = 1$ for the rest of $c_i^l \in C_{s_i}$.
\begin{itemize}
    \item If $f(c_i^l) = 1$, the weight of $\{b_i^l, c_i^l, g_i^l, h_i^l, p_i^l, q_i^l, w_i^l, x_i^l, y_i^l\} \cup Z_i^l$ will be three and $f(a_i) = 2$. 
    \item The weight of $\{b_i^l, c_i^l, g_i^l, h_i^l, p_i^l, q_i^l, w_i^l , x_i^l, y_i^l\} \cup Z_i^l$ will be three if $f(c_i^l) = 1$
\end{itemize}
    The total weight of $\{a_i\} \cup B_{s_i} \cup C_{s_i} \cup G_{s_i} \cup H_{s_i} \cup P_{s_i} \cup Q_{s_i} \cup W_{s_i} \cup X_{s_i} \cup Y_{s_i} \cup Z_{s_i}$ is $5+3\cdot (max(s_i)-1)$, which is $3\cdot max(s_i)+2$. 
        \medskip
    
\noindent \textbf{Case 5.} Let $f(c_i^l) = -1$ for some vertex $c_i^l \in C_{s_i}$.
\begin{itemize}
    \item The weight of $\{b_i^l, c_i^l, g_i^l, h_i^l, p_i^l, q_i^l, w_i^l , x_i^l, y_i^l\} \cup Z_i^l$ will be six and $f(a_i) = 2$.
    \item The weight of $\{b_i^l, c_i^l, g_i^l, h_i^l, p_i^l, q_i^l, w_i^l , x_i^l, y_i^l\} \cup Z_i^l$ will be three if $f(c_i^l) = 1$ or $f(c_i) = 2$.
\end{itemize}

    The total weight $\{a_i\} \cup B_{s_i} \cup C_{s_i} \cup G_{s_i} \cup H_{s_i} \cup P_{s_i} \cup Q_{s_i} \cup W_{s_i} \cup X_{s_i} \cup Y_{s_i} \cup Z_{s_i}$ is $6+3\cdot (max(s_i)-1)+2$, which is $3\cdot max(s_i)+5$.
\end{proof}
\begin{lemma}~\label{obs3}
       Let $f$ a minimum weighted signed Roman dominating function. For each $j \in \{1,2, ..., k\}$, $u_j$ is assigned the label 2 and $r_j^1$ is assigned the label $-1$ in $f$.
\end{lemma}
   \begin{proof}
       Consider a minimum weighted signed Roman dominating function $f$. There are two ways to label the vertex $u_j$ and its adjacent pendant vertex $r_j^1$ in $f$. 
           \medskip
    
\noindent \textbf{Case 1.} $u_j$ is labeled 2 and the adjacent pendant vertex $r_j^1$ is labeled $-1$.
    
\noindent \textbf{Case 2.} Both $u_j$ and $r_j^1$ are assigned the label 1.
    \medskip
    
In case 1, the weight of $\{u_j,r_j^1\}$ is one where as in case 2, the weight of $\{u_j,r_j^1\}$ is two. In both the cases the labelSum of $r_j^1$ is positive. In case 1, the labelSum of $u_j$ from the set $\{u_j,r_j^1\}$ is one where as in case 2, the labelSum of $u_j$ from the set $\{u_j,r_j^1\}$ is two. In case 1, we can swap the label of any neighbour of $u_j$ (excluding $r_j^1$) of label $1$ with label 2 to increase the labelSum of $u_j$ by one, with the same weight as that in case 2. From Lemma \ref{weightcases}, it is evident that the vertices of $C$ have labels from $\{1,2\}$ in minimum weighted $f$. Therfore, there always exists a neighbour for $u_j$ from $C$ with label 1, due to which the swap of a neighbour of $u_j$ from $C$ with label 1 to label 2 is always possible. Hence, we conclude that the case 1 is always superior to case 2. Therefore, we conclude that there exists a minimum weighted signed Roman dominating function $f$ with label 2 for $u_j$ and the label $-1$ for $r_j^1$. 
   \end{proof}
\begin{lemma}~\label{obs4}
    Let $f$ be a minimum weighted signed Roman dominating function. For each $j \in \{1,2, ..., k\}$, $f(v) = -1$ for all $v \in D_j$. 
\end{lemma}
\begin{proof}
From Lemma \ref{weightcases}, we have that, if $f(c_i^l)=1$ for each $c_i^l \in C_{s_i}$, the weight of $\{a_i\} \cup B_{s_i} \cup C_{s_i} \cup G_{s_i} \cup H_{s_i} \cup P_{s_i} \cup Q_{s_i} \cup W_{s_i} \cup X_{s_i} \cup Y_{s_i} \cup Z_{s_i}$ is $3 \cdot max(s_i)+1$. If $f(c_i^l)=2$ either for one vertex or subset of vertices of $C_{s_i}$, the weight of $f$ is $3 \cdot max(s_i)+2$. For each $s_i \in S$, the vertices in $C_{s_i}$ are adjacent to a subset of vertices in $\bigcup\limits_{j =1}^k \{u_j, v_j\}$. To obtain the minimum weight, the vertices in $C_{s_i}$ cannot be assigned the label $-1$. As each vertex in $C_{s_i}$ has labels from $\{0,1\}$, vertex $u_j$ has a positive labelSum from its neighbours in $\bigcup\limits_{s_i \in S}C_{s_i}$. Hence, we can assign the label $-1$ to a subset of vertices in $D_j$. Note that, even if we choose to label all the vertices of $D_j$ with $-1$, the labelSum of $v_j$ would still be positive. So, it only depends on $u_j$ regarding the number of vertices from $D_j$ that takes the label $-1$.
    
Now we argue that, every minimum weighted minus dominating function $f$ assigns the label $-1$, for each vertex in $D_j$, for all $j \in \{1,2,...,k\}$. For the sake of contradiction, let us assume that a subset of vertices in $D_j$ are labeled 0. Then, we select a vertex $v$ from $D_j$ with the label 1 and change its label from 1 to $-1$. Simultaneously, we select a vertex adjacent to $u_j$ in $C_{s_i}$ from some $s_i \in S$ and change its label from $-1$ to 1. After this transition in the labels, the labelSum of $u_j$ remains unchanged and the labelSum does not decrease for any vertex in the graph. Due to Lemma \ref{weightcases}, the weight of the set $A_{s_i} \cup B_{s_i} \cup C_{s_i}$ will either be $3 \cdot max(s_i) +1$ or $3 \cdot max(s_i) +2$. During this transition in the labels, the weight of $D_j$, for some $j \in \{1,2,...,k\}$, decreases by one, in each iteration. Whereas the weight of $\bigcup\limits_{s_i \in S} (A_{s_i} \cup B_{s_i} \cup C_{s_i}$), increases by one, at most $n$ times, once for each $s_i \in S$. For the first $n$ times during this transition, there may not be any decrease in the weight of $f$, but thereafter, the weight of $f$ decreases by one, in each iteration. Hence, after $n+c$ transitions, we obtain a function $f$ whose weight is reduced by $c$. In order to obtain the minimum weighted minus dominating function $f$, the transition has to happen until all the vertices in $D_j$ are labeled $-1$. Therefore, we conclude that, every minimum weighted minus dominating function $f$ assigns the label $-1$, for every vertex $v \in D_j$, for all $j \in \{1,2,...,k\}$.
\end{proof}
\begin{lemma}~\label{backward}
    \textit{If there exists a signed Roman dominating function of weight at most $k'$ then ($k, m, S, t$) is a yes instance of }MRSS problem.
\end{lemma} 
\begin{proof}
        Let $f$ be a signed Roman dominating function of weight $k' = \sum\limits_{s_i \in S}$ ($3 \cdot max(s_i)+1$) $-\sum\limits_{j \in \{1,2,...,k\}}$ $((\sum\limits_{s_i \in S} s_i(j))+t(j))$ $+2k+m$.
    \begin{itemize}
    \item Due to Lemma \ref{obs1}, Lemma \ref{obs2}, Lemma \ref{obs3} and Lemma \ref{obs4}, we fix the labels of $P_j = -1, F_j = 2, v_j = 2$ $, D_j = -1, u_j = 2$, $r_j^1 = -1$ and $r_j^2 = -1$ for each $j \in \{1,2,...,k\}$. 
    \item After assigning the labels to $\{u_j, v_j$, $r_j^1, r_j^2\}$ and to the vertices in sets $P_j, F_j$ and $D_j$, vertex $u_j$ has a negative labelSum of $-((\sum\limits_{s_i \in S} s_i(j))+t(j))+1$. For vertex $u_j$ to have a positive labelSum, we must set positive labels for its neighbours in $C$. The labelSum of its neighbours from $C$ must be at least $(\sum\limits_{s_i \in S} s_i(j))+t(j)$. 
    \item At this point, we are left with a weight of $\sum\limits_{s_i \in S} (3\cdot max(s_i)+1)+m$ for $\bigcup\limits_{s_i \in S} (\{a_i\} \cup B_{s_i} \cup C_{s_i} \cup G_{s_i} \cup H_{s_i} \cup P_{s_i} \cup Q_{s_i} \cup W_{s_i} \cup X_{s_i} \cup Y_{s_i} \cup Z_{s_i})$. From Lemma \ref{weightcases}, we have that, for some $s_i \in S$, if we label all the vertices in $C_{s_i}$ with 1, then the minimum weight of $\{a_i\} \cup B_{s_i} \cup C_{s_i} \cup G_{s_i} \cup H_{s_i} \cup P_{s_i} \cup Q_{s_i} \cup W_{s_i} \cup X_{s_i} \cup Y_{s_i} \cup Z_{s_i}$ is $3 \cdot max(s_i)+1$ and for some $s_i \in S$, if we label a subset of vertices in $C_{s_i}$ with $2$, then the minimum weight of $\{a_i\} \cup B_{s_i} \cup C_{s_i} \cup G_{s_i} \cup H_{s_i} \cup P_{s_i} \cup Q_{s_i} \cup W_{s_i} \cup X_{s_i} \cup Y_{s_i} \cup Z_{s_i}$ will be $3 \cdot max(s_i)+2$, needing an extra weight of one for each such set. 
    \item Based on the remaining weight of $\sum\limits_{s_i \in S} (3 \cdot max(s_i)+1)+m$ for $\{a_i\} \cup B_{s_i} \cup C_{s_i} \cup G_{s_i} \cup H_{s_i} \cup P_{s_i} \cup Q_{s_i} \cup W_{s_i} \cup X_{s_i} \cup Y_{s_i} \cup Z_{s_i}$, we can afford to label the vertices in $C_{s_i}$ with $2$ for at most $m$ such sets.
    \item Hence, we obtain that, if there exist $m$ sets from $C_{s_i}$ that bring positive labelSum to each vertex in $\bigcup_{j \in \{1,2,...,k\}}u_j$, then there exist $m$ vectors from $S$ that surpass $t$. 
    \end{itemize}
\end{proof}
From Lemma \ref{forward} and Lemma \ref{backward}, we arrive at the following theorem. 
\begin{theorem}
    SRD problem parameterized by feedback vertex set number is W[1]-hard.
\end{theorem} 
The parameters treewidth, clique-width are bounded by a function of the larger parameter feedback vertex set number. Hence, we obtain the following result for treewidth and clique-width.
\begin{corollary}~\label{cliquewidth}
    SRD problem is W[1]-hard parameterized by treewidth or clique-width.
\end{corollary} 
\section{FPT Parameterized by Neighbourhood Diversity}
From Corollary \ref{cliquewidth}, we have that SRD problem is W[1]-hard parameterized by clique-width. In this section, we study the parameterized complexity of SRD problem parameterized by neighbourhood diversity and present an FPT algorithm.

Lampis~\cite{lampis} proved that the neighbourhood diversity can be found in polynomial-time. For the rest of this section, we assume that the partition into $t$ sets $V_1, V_2, ..., V_t$ is given to us. It is to be noted that by the definition of neighbourhood diversity, each partition among $V_1, V_2, ..., V_t$ is either a clique or an independent set. 

In order to show the fixed-parameter tractability, we transform the problem into an instance of Integer Linear Programming (ILP) which is known to be FPT in the number of variables. The existing result related to ILP is given as follows.
\begin{theorem} [\cite{FLWS}] ~\label{theoremilp}
    The $p$-variable Integer Linear Programming problem can be solved using $\mathcal{O}(p^{2.5p+o(p)} \cdot L \cdot \log(MN))$ arithmetic operations and space polynomial in $L$, where $L$ is the number of bits in the input, $N$ is the maximum absolute value any variable can take, and $M$ is an upper bound on the absolute value of the minimum taken by the objective function.
\end{theorem} 
 \begin{itemize}
     \item Let $V_1, V_2, ..., V_t$ represent the partition of the vertex set $V$ into corresponding sets. We use $C$ to denote the union of cliques and $I$ to denote the union of independent sets among $V_1, V_2, ..., V_t$.
     \item Let $f$ be the minimum weighted signed Roman dominating function. We define three boolean variables $a_i$, $b_i$ and $c_i$ for each partition $V_i$. The variable $a_i$ indicates whether there exists a vertex in $V_i$ with the label $-1$ in $f$. Similarly, the variables $b_i$ and $c_i$ indicate whether there exists a vertex in $V_i$ with the label $1$ in $f$ and with the label $2$ in $f$, respectively. 
     \item We guess from each partition $V_i$ whether there exists a vertex $v$ with $f(v) = -1$. If true, we set $a_i = 1$ otherwise we set $a_i = 0$. We also guess from each partition whether there exists a vertex $v$ with $f(v) = 1$. If true, we set $b_i = 1$ otherwise we set $b_i = 0$. Similarly, we guess from each partition whether there exists a vertex $v$ with $f(v) = 2$. If true, we set $c_i = 1$ otherwise we set $c_i = 0$. For any partition $V_i$, if $a_i= 0, b_i = 0$ and $c_i =0$ then we simply reject the guess.
     \item For each partition $V_i \in I$ with $a_i=1$, we check whether there exists a partition $V_j \in N(V_i)$ with $c_j=1$. For each partition $V_i \in C$ with $a_i=1$, we check whether $c_i =1$ or there exists a partition $V_j \in N(V_i)$ with $c_j=1$. If for any partition, the above condition is not met, we reject the guess. The above conditions are to check whether there exists a neighbour with the label $2$ to a vertex in $V_i$ with the label $-1$.
     \item For each partition $V_i$, we define a variable $x_i$ that indicates the sum of the labels of all the vertices in $V_i$.
 \end{itemize}
Here, we transform the problem into an instance of ILP which is known to be FPT parameterized by the number of variables.
\medskip
    
\noindent The ILP formulation for the problem is given as follows.
\begin{tcolorbox}
{
    Minimize 
    $\sum\limits_{i| \{a_i\neq 1 \wedge b_i = 1 \wedge c_i=1\}} x_i$ + 
    $\sum\limits_{i| \{a_i = 1 \wedge b_i=1 \wedge c_i = 1\}} x_i$ + 
    $\sum\limits_{i| \{a_i= 1 \wedge b_i=1 \wedge c_i \neq 1 \wedge |V_i| \equiv 0 (mod 2)\}} 2x_i$ +
    $\sum\limits_{i| \{a_i= 1 \wedge b_i=1 \wedge c_i \neq 1 \wedge |V_i| \equiv 1 (mod 2)\}}$ $(2x_i+1)$ +
    $\sum\limits_{i| \{a_i= 1 \wedge b_i \neq1 \wedge c_i = 1 \wedge |V_i| \equiv 0 (mod 3)\}} $ $3x_i$ + 
    $\sum\limits_{i| \{a_i= 1 \wedge b_i \neq1 \wedge c_i = 1 \wedge |V_i| \equiv 1 (mod 3)\}} $ $(3x_i-1)$ + 
    $\sum\limits_{i| \{a_i= 1 \wedge b_i \neq1 \wedge c_i = 1 \wedge |V_i| \equiv 2 (mod 3)\}} $$(3x_i+1)$  
    \vspace{4mm}
    
    \noindent Subject to
    \begin{enumerate}
    \item[(C1)] $x_i + A \geq 1$ for each $V_i \in C$. 
    \item[(C2)] $-1 + A \geq 1$ for each $V_i \in I$ and $a_i = 1$.
    \item[(C3)] $1 + A \geq 1$ for each $V_i \in I$ and $a_i \neq 1$ and $b_i= 1$. 
    \item[(C4)] $2 + A \geq 1$ for each $V_i \in I$ and $a_i \neq 1$, $b_i \neq 1$ and $c_i = 1$.
    \item[(C5)] $|V_i|+1 \leq x_i \leq 2|V_i|-1$, for each $V_i$ with $a_i \neq 1$, $b_i = 1$ and $c_i =1.$
    \item[(C6)] $-|V_i|+5 \leq x_i \leq 2|V_i|-4$, for each $V_i$ with $a_i = 1, b_i = 1$ and $c_i =1.$
    \item[(C7)] $\lceil \frac{-|V_i|}{2}\rceil +1\leq x_i \leq \lfloor \frac{|V_i|}{2} \rfloor -1$, for each $V_i$ with $a_i = 1, b_i = 1$ and $c_i \neq 1$.
    \item[(C8)] $\lfloor \frac{-|V_i|+1}{3}\rfloor+1 \leq x_i \leq \lceil \frac{|V_i|}{3} \rceil +1$, for each $V_i$ with $a_i = 1, b_i \neq 1$ and $c_i =1.$
    \end{enumerate}
    Where $A$ =  $\sum\limits_{j| \{V_j \in N(V_i) \wedge a_j\neq 1 \wedge b_j = 1 \wedge c_j=1\}} x_i$ + 
    $\sum\limits_{j| \{V_j \in N(V_i) \wedge a_j = 1 \wedge b_i=1 \wedge c_j = 1\}} x_j$ + 
    $\sum\limits_{j| \{V_j \in N(V_i) \wedge a_j= 1 \wedge b_j=1 \wedge c_j \neq 1 \wedge |V_j| \equiv 0 (mod 2)\}} 2x_j$ +
    $\sum\limits_{j| \{V_j \in N(V_i) \wedge a_j= 1 \wedge b_j=1 \wedge c_j \neq 1 \wedge |V_j| \equiv 1 (mod 2)\}}$ $(2x_j+1)$ +
    $\sum\limits_{j| \{V_j \in N(V_i) \wedge a_j= 1 \wedge b_j \neq1 \wedge c_j = 1 \wedge |V_j| \equiv 0 (mod 3)\}} $ $3x_j$ + 
    $\sum\limits_{j| \{V_j \in N(V_i) \wedge a_j= 1 \wedge b_j \neq1 \wedge c_j = 1 \wedge |V_j| \equiv 1 (mod 3)\}} $ $(3x_j-1)$ + 
    $\sum\limits_{j| \{V_j \in N(V_i) \wedge a_j= 1 \wedge b_j \neq1 \wedge c_j = 1 \wedge |V_j| \equiv 2 (mod 3)\}} $$(3x_j+1)$
}
\end{tcolorbox} 
\noindent We now discuss the constraints given in the ILP formulation. 
    \medskip
    
\noindent (C1) For each partition $V_i$ that is a clique, sum of the labels of $V_i$ and the labels of its neighbouring partitions must be at least one. 
    \medskip
    
\noindent (C2) For each partition $V_i$ that is an independent set, if $a_i = 1$, then there exists a vertex in $V_i^k$ with label $-1$.  
In order for $V_i^k$ to have a positive labelSum, the value $-1$ plus the sum of the labels of its neighboring partitions must be at least one. 
    \medskip
    
\noindent (C3) For each partition $V_i$ that is a independent set, if $a_i \neq 1$ and $b_i = 1$, then there exists some vertex in $V_i^k$ with the label 1 and no vertex of label $-1$ exists. In order for $V_i^k$ to have a positive labelSum, the value $1$ plus the sum of the labels of its neighbouring partitions must be at least one. 
    \medskip
    
\noindent (C4) For each partition $V_i$ that is a independent set, if $a_i \neq 1$, $b_i \neq 1$ and $c_i = 1$, then there exists some vertex in $V_i^k$ with the label 2 and no vertex from labels $\{-1, 1\}$ exists. In order for $V_i^k$ to have a positive labelSum, the value $2$ plus the sum of the labels of its neighbouring partitions must be at least one. 
    \medskip
    
\noindent (C5) Consider a partition $V_i$ with $a_i \neq 1$, $b_i = 1$ and $c_i =1$. The value of $x_i$ is minimum if there are $|V_i|-1$ vertices with the label $1$ and one vertex with the label $2$. Therefore, the lower bound for $x_i$ is $|V_i|+1$. The value of $x_i$ is maximum if there are $|V_i|-1$ vertices with the label $2$ and one vertex with the label $1$. Hence, the upper bound is $2|V_i|-1$.
    \medskip
    
\noindent (C6) Consider a partition $V_i$ with $a_i = 1, b_i = 1$ and $c_i =1$. The value of $x_i$ is minimum if there are $|V_i|-2$ vertices with the label $-1$ and one vertex each with the label $1$ and $2$. Therefore, the lower bound for $x_i$ is $-|V_i|+5$. The value of $x_i$ is maximum if there are $|V_i|-2$ vertices with the label $2$ and one vertex each with the label $1$ and $-1$. Hence, the upper bound is $2|V_i|-4$.
    \medskip
    
\noindent (C7) Consider a partition $V_i$ with $a_i = 1, b_i = 1$ and $c_i \neq 1$. The value of $x_i$ is minimum if there are $|V_i|-1$ vertices with the label $-1$ and one vertex with the label $1$. Therefore, the lower bound for $x_i$ is ${-|V_i|} +2$. The value of $x_i$ is maximum if there are $|V_i|-1$ vertices with the label $1$ and one vertex with the label $-1$. Hence, the upper bound is ${|V_i|} -2$. If $|V_i|$ is odd then $x_i$ could take only odd values between ${-|V_i|} +2$ and ${|V_i|} -2$. Thus, $x_i$ $ \in [2 \cdot (\lceil \frac{|V_i|}{2} \rceil +1)+1, 2 \cdot (\lfloor \frac{|V_i|}{2} \rfloor -1)+1]$. Whereas if $|V_i|$ is even then $x_i$ could take only even values between ${-|V_i|} +2$ and ${|V_i|} -2$. Thus, $x_i$ $ \in [2 \cdot (\lceil \frac{|V_i|}{2} \rceil +1), 2 \cdot (\lfloor \frac{|V_i|}{2} \rfloor -1)]$.
    \medskip
    
\noindent (C8) Consider a partition $V_i$ with $a_i = 1, b_i \neq 1$ and $c_i =1$. The value of $x_i$ is minimum if there are $|V_i|-1$ vertices with the label $-1$ and one vertex with the label $2$. Therefore, the lower bound for $x_i$ is ${-|V_i|} +3$. The value of $x_i$ is maximum if there are $|V_i|-1$ vertices with the label $2$ and one vertex with the label $-1$. Hence, the upper bound is ${2\cdot|V_i|} -3$. If $|V_i|$ is $0 \pmod 3$, we have that $x_i$ $ \in [3 \cdot (\lfloor \frac{-|V_i|+1}{3}\rfloor+1), 3 \cdot (\lceil \frac{|V_i|}{3} \rceil +1)]$. Whereas if $|V_i|$ is $1 \pmod 3$, we have that $x_i$ $ \in [3 \cdot (\lfloor \frac{-|V_i|+1}{3}\rfloor+1)-1, 3 \cdot (\lceil \frac{|V_i|}{3} \rceil +1)-1]$. Similarly, if $|V_i|$ is $2 \pmod 3$, we have that $x_i$ $ \in [3 \cdot (\lfloor \frac{-|V_i|+1}{3}\rfloor+1)+1, 3 \cdot (\lceil \frac{|V_i|}{3} \rceil +1)+1]$. 
    \medskip
    

Now we discuss on how we assign labels to the vertices for the vertices in each partition $V_i$. We consider the guess on the values of $a_i$, $b_i$ and $c_i$, for the partition $V_i$. 
\begin{itemize}
    \item For each partition $V_i$ with $a_i \neq 1 $, $ b_i \neq 1$ and $c_i = 1$, we assign the label 2 to each vertex of $V_i$ in $f$. 
    \item For each partition $V_i$ with $a_i \neq 1$, $b_i = 1$ and $ c_i \neq 1$, we assign the label 1 to each vertex of $V_i$ in $f$. 
    \item For each partition $V_i$ with $a_i = 1, b_i \neq 1 $ and $ c_i \neq 1$, we assign the label $-1$ to each vertex of $V_i$ in $f$. 
\end{itemize}
    For all the other partitions, we assign the labels based on the outcome of ILP formulation. 
\begin{itemize}
    \item For each partition $V_i$ with $a_i \neq 1 $, $ b_i = 1$ and $c_i = 1$, we label each vertex of $V_i$ from $\{1, 2\}$ such that the weight of $V_i$ is exactly $x_i$ and there must exist two vertices $u$ and $v$ in the partition such that $f(u) = 1$ and $f(v) = 2$. 
    \item For each partition $V_i$ with $a_i = 1 $, $ b_i = 1$ and $c_i = 1$, we label each vertex of $V_i$ from $\{-1, 1, 2\}$ such that the weight of $V_i$ is exactly $x_i$ and there must exist three vertices $u,v$ and $w$ in the partition such that $f(u) = -1$, $f(v) = 1$ and $f(w) = 2$. 
    \item For each partition $V_i$ with $a_i = 1 $, $ b_i = 1$ and $c_i \neq 1$, we label each vertex of $V_i$ from $\{-1, 1\}$ such that the weight of $V_i$ is exactly $x_i$ and there must exist two vertices $u$ and $v$ in the partition such that $f(v) = -1$ and $f(v) = 1$. 
    \item Similarly, for each partition $V_i$ with $a_i = 1 $, $ b_i \neq 1$ and $c_i = 1$, we label each vertex of $V_i$ from $\{-1, 2\}$ such that the weight of $V_i$ is exactly $x_i$ and there must exist two vertices $u$ and $v$ in the partition such that $f(v) = -1$ and $f(v) = 2$. 
\end{itemize}

    In this way, we label the vertices of the partitions $V_1, V_2, ..., V_t$.

In our ILP formulation, we have $t$ variables. The values of all the variables and the objective function are upper bounded by $n$. With the help of Theorem \ref{theoremilp}, we will be able to solve the problem in $\mathcal{O}^*(8^t \cdot t^{\mathcal{O}(t)})$ time.

Hence, we arrive at the following theorem. 
\begin{theorem}
    SRD problem is fixed-parameter tractable parameterized by neighbourhood diversity.
\end{theorem} 
As the parameter neighbourhood diversity is bounded by a function of the larger parameter vertex cover number, we obtain the following result.
\begin{corollary}~\label{corollaryvc}
    SRD problem is fixed-parameter tractable parameterized by vertex cover number.
\end{corollary}
\section{No Polynomial Kernel Parameterized by Vertex Cover Number}
From Corollary \ref{corollaryvc}, we have that SRD problem is FPT parameterized by vertex cover number. In this section, we complement this result by proving that the problem does not admit a polynomial kernel parameterized by vertex cover number unless coNP $\subseteq$ NP/poly.

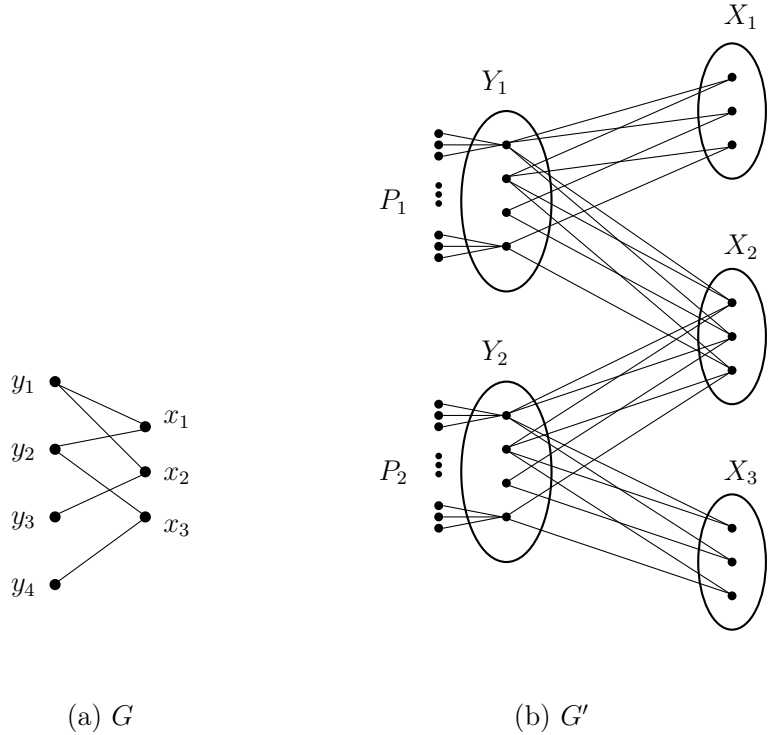
\begin{figure} [t]
\centering
    \begin{tikzpicture} [thick,scale=0.6, every node/.style={scale=0.9}]
        \draw (5, 8) ellipse (1 and 2);
        \draw (5, 2) ellipse (1 and 2);
        
        \draw (10, 10) ellipse (0.75 and 1.5);
        \draw (10, 5) ellipse (0.75 and 1.5);
        \draw (10, 0) ellipse (0.75 and 1.5);

        \filldraw (5, 9.25) circle (0.075cm);
        \filldraw (5, 8.5) circle (0.075cm);
        \filldraw (5, 7.75) circle (0.075cm);
        \filldraw (5, 7) circle (0.075cm);

        \filldraw (5, 3.25) circle (0.075cm);
        \filldraw (5, 2.5) circle (0.075cm);
        \filldraw (5, 1.75) circle (0.075cm);
        \filldraw (5, 1) circle (0.075cm);

        \filldraw (3.5, 3.25) circle (0.075cm);
        \filldraw (3.5, 3) circle (0.075cm);
        \filldraw (3.5, 3.5) circle (0.075cm);

        \filldraw (3.5, 2.35) circle (0.05cm);
        \filldraw (3.5, 2.15) circle (0.05cm);
        \filldraw (3.5, 1.95) circle (0.05cm);

        \filldraw (3.5, 1) circle (0.075cm);
        \filldraw (3.5, 1.25) circle (0.075cm);
        \filldraw (3.5, 0.75) circle (0.075cm);

        \filldraw (3.5, 9.25) circle (0.075cm);
        \filldraw (3.5, 9.5) circle (0.075cm);
        \filldraw (3.5, 9) circle (0.075cm);

        \filldraw (3.5, 8.35) circle (0.05cm);
        \filldraw (3.5, 8.15) circle (0.05cm);
        \filldraw (3.5, 7.95) circle (0.05cm);

        \filldraw (3.5, 7) circle (0.075cm);
        \filldraw (3.5, 7.25) circle (0.075cm);
        \filldraw (3.5, 6.75) circle (0.075cm);

        \draw[thin] (3.6, 3.25) -- (4.9, 3.25);
        \draw[thin] (3.6, 3) -- (4.9, 3.25);
        \draw[thin] (3.6, 3.5) -- (4.9, 3.25);


        
        \draw[thin] (3.6, 1) -- (4.9, 1);
        \draw[thin] (3.6, 0.75) -- (4.9, 1);
        \draw[thin] (3.6, 1.25) -- (4.9, 1);

        \draw[thin] (3.6, 9.25) -- (4.9, 9.25);
        \draw[thin] (3.6, 9) -- (4.9, 9.25);
        \draw[thin] (3.6, 9.5) -- (4.9, 9.25);


        
        \draw[thin] (3.6, 7) -- (4.9, 7);
        \draw[thin] (3.6, 7.25) -- (4.9, 7);
        \draw[thin] (3.6, 6.75) -- (4.9, 7);

        \filldraw (10, 9.25) circle (0.075cm);
        \filldraw (10, 10) circle (0.075cm);
        \filldraw (10, 10.75) circle (0.075cm);

        \filldraw (10, 4.25) circle (0.075cm);
        \filldraw (10, 5) circle (0.075cm);
        \filldraw (10, 5.75) circle (0.075cm);

        \filldraw (10, -0.75) circle (0.075cm);
        \filldraw (10, 0) circle (0.075cm);
        \filldraw (10, 0.75) circle (0.075cm);

        \draw[thin] (9.95, 9.95) -- (5.05, 7.8);

        \draw[thin] (9.95, 9.2) -- (5.05, 7.05);

        \draw[thin] (9.95, 9.2) -- (5.05, 8.55);
        \draw[thin] (9.95, 10.7) -- (5.05, 8.55);

        \draw[thin] (9.95, 9.95) -- (5.05, 9.3);
        \draw[thin] (9.95, 10.7) -- (5.05, 9.3);

        \draw[thin] (9.95, 5.05) -- (5.05, 7.7);

        \draw[thin] (9.95, 4.3) -- (5.05, 6.95);

        \draw[thin] (9.95, 4.3) -- (5.05, 8.45);
        \draw[thin] (9.95, 5.8) -- (5.05, 8.45);

        \draw[thin] (9.95, 5.05) -- (5.05, 9.2);
        \draw[thin] (9.95, 5.8) -- (5.05, 9.2);

        \draw[thin] (9.95, 4.95) -- (5.05, 3.3);
        \draw[thin] (9.95, 5.7) -- (5.05, 3.3);

        \draw[thin] (9.95, 4.2) -- (5.05, 2.55);
        \draw[thin] (9.95, 5.7) -- (5.05, 2.55);

        \draw[thin] (9.95, 4.95) -- (5.05, 1.8);

        \draw[thin] (9.95, 4.2) -- (5.05, 1.05);

        \draw[thin] (9.95, 0.05) -- (5.05, 3.2);
        \draw[thin] (9.95, 0.8) -- (5.05, 3.2);

        \draw[thin] (9.95, -0.7) -- (5.05, 2.45);
        \draw[thin] (9.95, 0.8) -- (5.05, 2.45);

        \draw[thin] (9.95, 0.05) -- (5.05, 1.7);

        \draw[thin] (9.95, -0.7) -- (5.05, 0.95);

        \draw[thin] (-4.93, 3.93) -- (-3.07, 3.07);
        \draw[thin] (-4.93, 2.57) -- (-3.07, 2.93);
        \draw[thin] (-4.93, 3.93) -- (-3.07, 2.07);
        \draw[thin] (-4.93, 2.43) -- (-3.07, 1.07);
        \draw[thin] (-4.93, 1.07) -- (-3.07, 1.93);
        \draw[thin] (-4.93, -0.43) -- (-3.07, 0.93);

        \filldraw (-5, 4) circle (0.1cm);
        \filldraw (-5, 2.5) circle (0.1cm);
        \filldraw (-5, 1) circle (0.1cm);
        \filldraw (-5, -0.5) circle (0.1cm);
        
        \filldraw (-3, 3) circle (0.1cm);
        \filldraw (-3, 2) circle (0.1cm);
        \filldraw (-3, 1) circle (0.1cm);

        \draw (-2.3, 0.35) circle (0cm) node[anchor=south]{$x_3$};
        \draw (-2.3, 1.5) circle (0cm) node[anchor=south]{$x_2$};
        \draw (-2.3, 2.75) circle (0cm) node[anchor=south]{$x_1$};

        \draw (-5.7, 3.5) circle (0cm) node[anchor=south]{$y_1$};
        \draw (-5.7, 2) circle (0cm) node[anchor=south]{$y_2$};
        \draw (-5.7, 0.5) circle (0cm) node[anchor=south]{$y_3$};
        \draw (-5.7, -1) circle (0cm) node[anchor=south]{$y_4$};

        \draw (2.5, 7.5) circle (0cm) node[anchor=south]{$P_1$};


        \draw (2.5, 1.5) circle (0cm) node[anchor=south]{$P_2$};


        \draw (4.75, 4.15) circle (0cm) node[anchor=south]{$Y_2$};
        \draw (4.75, 10.15) circle (0cm) node[anchor=south]{$Y_1$};
        
        \draw (10.2, 1.5) circle (0cm) node[anchor=south]{$X_3$};
        \draw (10.2, 6.5) circle (0cm) node[anchor=south]{$X_2$};
        \draw (10.2, 11.6) circle (0cm) node[anchor=south]{$X_1$};

        \draw (-4, -4) circle (0cm) node[anchor=south]{(a) $G$};

        \draw (6, -4) circle (0cm) node[anchor=south]{(b) $G'$};

    \end{tikzpicture}
    \caption{(a) a graph $G$ with $k$ = 2 and (b) the graph $G'$ with $k'$ = $-3$}
    \label{fig: Figvc}
\end{figure}

We provide a polynomial parameter transformation (PPT) from the well-known \rbds{} (RBDS) problem to prove that SRD problem does not admit a polynomial kernel parameterized by vertex cover number unless coNP $\subseteq$ NP/poly. In this regard, we define the concept of polynomial parameter transformation. 
\begin{definition}
    Consider two parameterized languages $P$ and $Q$. There exists a polynomial parameter transformation from $P$ to $Q$ if there is a polynomial time computable function $f :\sum^* \times\mathbb{N} \rightarrow \sum^* \times\mathbb{N}$, a polynomial $p : \mathbb{N} \rightarrow \mathbb{N}$ such that $(x,k)  \in P$ if and only if $f(x,k) \in Q $ and $k' \leq p(k)$ where $f((x,k))= (x',k')$. 
\end{definition}
The problem RBDS is defined as follows.
\begin{tcolorbox}
{
\textbf{Problem.} \rbds{} \newline
\textbf{Input.} A bipartite graph $G = (X \cup Y, E)$ and an integer $k$\newline
\textbf{Output.} Does there exist a subset $S \subseteq X$ of size at most $k$ that dominates $Y$?
}
\end{tcolorbox}
The following result is from the literature.
\begin{theorem}[\cite{kernel}]
     RBDS problem parameterized by $|Y|$ does not admit a polynomial kernel unless coNP $\subseteq$ NP/poly.
\end{theorem} 
\noindent \textbf{Construction.} Consider an instance $I = (G, k)$ of RBDS problem, we construct the corresponding SRD problem instance $I' = (G', k')$ as follows. 
\begin{itemize}
    \item We create three distinct copies $X_1, X_2$ and $X_3$ of $X$ and two distinct copies $Y_1$ and $Y_2$ of $Y$ in $G'$.
    \item The vertex corresponding to $u \in X$ is denoted by $X_1^u, X_2^u$ and $X_3^u$ in $X_1, X_2$ and $X_3$ respectively. Similarly, the vertex corresponding to $u \in Y$ is denoted by $Y_1^u$ and $Y_2^u$ in $Y_1$ and $Y_2$ respectively.
    \item We make each vertex $u \in Y_1$ adjacent to the vertices, $X_1^{N_G(u)}$ in $X_1$ and $X_2^{N_G(u)}$ in $X_2$. We also make each vertex $u \in Y_2$ adjacent to the vertices, $X_2^{N_G(u)}$ in $X_2$ and $X_3^{N_G(u)}$ in $X_3$.
    \item We add three pendant vertices to each vertex of the sets $Y_1$ and $Y_2$. The pendant vertices adjacent to the vertices of the sets $Y_1$ and $Y_2$ are denoted by the sets $P_1$ and $P_2$, respectively.
\end{itemize}    This completes the construction of the reduced instance $G'$. For more details, see \figureautorefname~\ref{fig: Figvc} for an illustration.

We set $k' = -2|Y|-|X|+4k$ and the set corresponding to vertex cover number is $Y_1 \cup Y_2$ of size $2|Y|$.

\begin{lemma}
    There exists a subset $S \subseteq X$ of size at most $k$ that dominates $Y$ if and only if there exists a signed Roman dominating function $f$ of weight at most $k'$.
\end{lemma}
\begin{proof}
    $(\Rightarrow)$ Let $S$ be a subset of $X$ that dominates $Y$. The corresponding signed Roman dominating function $f$ is given as follows.
            \[f(w) =\begin{cases}
            -1, & \text{if } w \in \left(P_1 \cup P_2 \cup \bigcup\limits_{v \notin S} (X_1^v \cup X_3^v)\right), \\
    1, & \text{if } w \in \left(X_2 \cup \bigcup\limits_{v \in S} (X_1^v \cup X_3^v)\right), \\
    2, & \text{if } w \in \left(Y_1 \cup Y_2\right)
        \end{cases}\]
\begin{itemize}
    \item The vertices in the sets $Y_1$ and $Y_2$ gets a positive labelSum from their adjacent vertices in $X_1 \cup X_2 \cup X_3$.
    \item The vertices in the sets $X_1, X_2$ and $X_3$ also has a positive labelSum due to the positive labels of their neighbours in $Y_1 \cup Y_2$.
    \item The pendant vertices in the sets $P_1$ and $P_2$ have a positive labelSum due to the positive labels of their neighbours in $Y_1 \cup Y_2$.
\end{itemize} 
Each vertex $u \in V(G')$ has a positive labelSum and the weight of the signed Roman dominating function is $k' = -2|Y|-|X|+4k$. 
    \medskip
    
\noindent $(\Leftarrow)$ Let $f$ be a signed Roman dominating function of weight $k' = -2|Y|-|X|+4k$. 
\begin{itemize}
    \item 
The pendant vertices $P_1 \cup P_2$ must get the label $-1$ in $f$ and their neighbours in $Y$ must get the label $2$ in $f$. Hence, the total weight of $P_1 \cup P_2 \cup Y_1 \cup Y_2$ is $-2|Y|$. 
\item The weight of the vertex set $X_1, X_2$ and $X_3$ combined must be at most $-|X|+4k$. The only way to achieve such a weight is to have the label $-1$ for majority of the vertices in $X_1$, $X_2$ and $X_3$. 
\item  We can assign the label $-1$ to each vertex in $X_1 \cup X_2 \cup X_3$, as the subgraph induced on the vertex set $X_1 \cup X_2 \cup X_3$ is an independent set and the labelSum of each vertex in $X_1 \cup X_2 \cup X_3$ remains at least one. But, we have to label a subset of vertices from $X_1 \cup X_2 \cup X_3$ with $-1$, in a way that, each vertex in $Y_1 \cup Y_2$ will have a positive labelSum.
\item The sum of the labels of each vertex in $Y_1 \cup Y_2$ from its neighbours in $X_1 \cup X_2 \cup X_3 \cup Y_1 \cup Y_2$ must be at least two. Therefore, we must label some vertices in $X_1 \cup X_2 \cup X_3$ with $1$.
\item A subset of the vertices in $X_1 \cup X_2 \cup X_3$ receives a label from $\{1,2\}$, while the majority are assigned the label $-1$. Since $X_2$ is adjacent to both $Y_1$ and $Y_2$, we label all the vertices in $X_2$ with 2. We label all vertices in $X_1$ and $X_3$ with $-1$, except $k$ vertices in each of them, which are assigned the label 1. These $k$ vertices with label 1 in $X_1$ and $X_3$ correspond to the $k$-sized subset of $X$ that dominates $Y$ in the RBDS instance.
\end{itemize}
\end{proof}
With this, we arrive at the following result.
\begin{theorem}
    SRD problem does not admit a polynomial kernel parameterized by vertex cover number unless coNP $\subseteq$ NP/poly.
\end{theorem}
\section{Conclusion}
\noindent In this paper, we investigated the complexity aspects of SRD problem. We proved that the problem is NP-complete on split graphs. We also showed that the problem is W[2]-hard parameterized by weight on bipartite graphs and W[1]-hard parameterized by weight on circle graphs. Further, we proved that the problem is W[1]-hard parameterized by feedback vertex set number (and hence for treewidth), whereas SD problem and RD problem admit FPT algorithms for treewidth. This result differentiates the parameterized complexity of SRD problem from that of SD problem and RD problem. Later, we presented an FPT algorithm parameterized by neighbourhood diversity. In addition, we proved that the problem does not admit a polynomial kernel parameterized by vertex cover number unless coNP $\subseteq$ NP/poly. 
\medskip

\noindent We conclude the paper with the following open questions. 
\begin{enumerate}
    \item What is the complexity of SRD problem on trees, cographs, block graphs and interval graphs?
    \item What is the parameterized complexity of SRD problem for various structural parameters such as distance to disjoint paths, pathwidth, twin cover number, feedback edge set number, max-leaf number, distance to cluster and modular-width?
    \item What is the parameterized complexity of SRD problem parameterized by weight on split graphs and chordal bipartite graphs?
    \item Does there exists a polynomial kernel for the SRD problem parameterized by distance to clique, neighbourhood diversity and feedback edge set number?
\end{enumerate}



\bibliographystyle{elsarticle-num}
\bibliography{elsarticle/references}
\end{document}